\documentclass[a4paper,oneside,12pt]{article}

\usepackage{epsfig,array,amsmath,amssymb,psfrag,graphicx,tensor,color,mathrsfs}

\definecolor{dblue}{RGB}{0, 0, 150}
\definecolor{color2}{RGB}{0,140,0}

\usepackage[colorlinks=true, linkcolor=blue, citecolor=color2, urlcolor=dblue]{hyperref}

\newcommand{\noi}{\noindent}
\newcommand{\beq}{\begin{equation}}
\newcommand{\eeq}{\end{equation}}
\newcommand{\bea}{\begin{eqnarray}}
\newcommand{\eea}{\end{eqnarray}}

\newcommand{\RR}{\mathbb{R}}

\newcommand{\mc}[1]{\mathcal{#1}}

\newcommand{\mr}[1]{\mathrm{#1}}

\newcommand{\cF}{\mathcal{F}}

\newcommand{\Tc}{\hat{T}}

\newcommand{\qqquad}{\quad\qquad}

\newtheorem{prop}{Proposition}[section]
\newtheorem{lemma}{Lemma}

\makeatletter \@addtoreset{equation}{section} \makeatother 
\renewcommand{\theequation}{\arabic{section}.\arabic{equation}}

\begin{document}

\title{\bf Energy-momentum tensor and duality symmetry of linearized gravity\\
in the Fierz formalism}

\author{G\'abor Zsolt T\'oth\\[4mm] 
\small \textit{Wigner Research Centre for Physics,} \\
\small \textit{Konkoly-Thege Mikl\'os \'ut 29-33,} \\
\small \textit{1121 Budapest, Hungary} \\
\small \texttt{toth.gabor.zsolt@wigner.hu}}
\date{}

\maketitle

\begin{abstract}
A formulation of linearized gravity in flat background,
based on the Fierz tensor as a counterpart of the electromagnetic field strength,
is discussed in detail and used to study fundamental properties of the linearized gravitational field.
In particular, the linearized Einstein equations are written as first order partial differential equations in terms of the Fierz tensor,
in analogy with the first order Maxwell equations.
An energy-momentum tensor ($T_{\mathrm{lg}}^{ab}$)
with favourable properties and exhibiting remarkable similarity 
to the standard energy-momentum tensor of the electromagnetic field
is found for the linearized gravitational field.
$T_{\mathrm{lg}}^{ab}$ is quadratic in the Fierz tensor
(which is constructed from the first derivatives of the linearized metric), traceless,
and satisfies the dominant energy condition
in a gauge that contains the transverse traceless gauge.
It is further shown that in suitable gauges,
including the transverse traceless gauge, 
linearized gravity in the absence of matter
has a duality symmetry that maps the Fierz tensor,
which is antisymmetric in its first two indices, into its dual.
Conserved currents associated with the gauge and duality symmetries
of linearized gravity are also determined.
These currents show good analogy with the corresponding currents in electrodynamics.
\end{abstract}

\noi

\thispagestyle{empty}

\newpage

%\tableofcontents

%\newpage

\section{Introduction}
\label{sec.intr}

Correspondences between general relativity and electrodynamics were noticed long ago
and have been investigated and applied extensively
throughout the past decades (see \cite{Matte}-\cite{UmGal3} and references therein).
Such correspondences become especially apparent if weak gravitational fields in flat background are considered,
as in this limit the gravitational field becomes a special relativistic massless field with an abelian gauge symmetry.

Linearized gravity in flat background has a straightforward but relatively little used formulation
that is Lorentz covariant, does not require gauge fixing or
field variables that contain higher derivatives of the linearized metric tensor
than the usual formulation,
and makes linearized gravity appear quite similar to electrodynamics.
The aim of this paper is to discuss this
formulation---which will be called Fierz formalism---and
to apply it to study fundamental
dynamical properties of the linearized gravitational field.
In particular, an energy-momentum tensor
that is analogous to the energy-momentum tensor of the electromagnetic
field and has favourable properties is found,
it is shown that linearized gravity in the absence of matter has a duality symmetry suited to the Fierz formalism,
and conserved currents associated with the duality and gauge symmetries
of linearized gravity are determined.
These investigations are also intended to further develop the Fierz formalism and to demonstrate its usefulness.

The Fierz formalism
originates from studies of fields of arbitrary spin by Fierz and Pauli \cite{Fierz,FP2}.
It rests on the Fierz tensor $F_{abc}$ (see (\ref{eq.lg.x6})),
which is constructed from the first derivatives of the linearized metric tensor $h_{ab}$
and is analogous, to some extent, to the electromagnetic field strength tensor $F_{ab}$. 
Although $F_{abc}$ is not completely gauge invariant, it is invariant under the action of those gauge transformations
that are of the form $h_{ab}\to h_{ab}+\partial_{ab}\phi$, where $\phi$ is an arbitrary function.
By adding a total divergence to the usual Fierz--Pauli Lagrangian of linearized gravity, it is possible
to obtain a Lagrangian that can be expressed in terms of $F_{abc}$ and is very similar to
the usual Lagrangian of the electromagnetic field. The linearized Einstein equations also take a form
in terms of $F_{abc}$ that closely resembles the inhomogeneous Maxwell equations,
and a counterpart of the homogeneous Maxwell equations exists as well.

Regarding the correspondence between the Fierz tensor and $F_{ab}$,
it must be noted that the most appropriate generalization of $F_{ab}$ is the linearized Riemann tensor
in the case of spin-2 fields; from this point of view, $F_{abc}$ is only a `semi-analogue' of $F_{ab}$.
The analogy established by the Fierz formalism between linearized gravity and electrodynamics 
has to be understood with this caveat in mind.
Nevertheless, in favour of $F_{abc}$ we recall that
the linearized Einstein equations expressed in terms of the linearized Riemann tensor
are not similar to Maxwell's equations, and the same can be said about the Lagrangians.

Formally the Fierz tensor is closely related to the Lanczos potential \cite{Lanczos}-\cite{JWK},
nevertheless the role of the Lanczos potential is different from that of the Fierz tensor;
the Lanczos potential acts as a potential for the Weyl tensor,
whereas the Fierz tensor is a field strength with $h_{ab}$ as its potential.

The problem of finding a suitable definition of the energy and momentum of the gravitational field in
general relativity has received very much attention and has vast literature \cite{Szabados}.
It seems that the gravitational field does not have local energy and momentum currents that have all the
desirable properties one would expect, and
this conclusion extends to the linearized gravitational field as well. 
On the other hand, there are many possibilities under less strict conditions
(for energy-momentum pseudotensors, in particular, see \cite{Szabados}-\cite{Petrov}; about linearized gravity,
see \cite{BHL1, Bicak, Bicak2, BS, Baker}). 
In view of this situation, it appears reasonable to search for special energy-momentum tensors that have at least some favourable properties,
in addition to the basic properties of Lorentz covariance, locality,
conservation in the absence of matter, and having a connection with spacetime translations.
It can be expected that in this way a not very large number of outstanding
energy-momentum tensors, having their own advantages, will be identified.
Ideally, a single optimal energy-momentum tensor might emerge.

The Fierz formalism opens the possibility of looking for an energy-momentum tensor
that fits the Fierz formalism and is similar to the energy-momentum tensor $T_{\mathrm{em}}^{ab}$ of the electromagnetic field;
in this paper we look for such an energy-momentum tensor.
We find that an energy-momentum tensor of this kind indeed exists
(it will be denoted by $T_{\mathrm{lg}}^{ab}$),
and has several favourable properties.
In obtaining $T_{\mathrm{lg}}^{ab}$,
as well as the conserved currents associated with the gauge and duality symmetries,
we apply Noether's first theorem in a modern form, combined with additional considerations.

In order to provide a broader perspective and to compare $T_{\mathrm{lg}}^{ab}$ with other energy-momentum tensors
that have appeared in the literature,
we also recall and further develop, making use of the Fierz formalism,
certain remarkable recent results \cite{Barnett, BHL1, BHL2, BS} (see also \cite{Butcher}).

In \cite{BHL1}, Butcher, Hobson and Lasenby proposed an energy-momentum tensor $\tau^{ab}$
in harmonic gauge, which is symmetric,
a homogeneous quadratic expression in $\partial_c h_{ab}$,
and satisfies the dominant energy condition\footnote{Although the term
``dominant energy condition'' is applied usually to the energy-momentum tensor of matter,
in this paper we use it in relation to the energy-momentum tensor of the linearized gravitational field as well.}
(which will be abbreviated as d.e.c.)
if $h_{ab}$ is also transverse and traceless.
In \cite{BHL2}, further results on $\tau^{ab}$ were presented, including a derivation of $\tau^{ab}$ using variational techniques
(which were avoided in \cite{BHL1}) and a generalization beyond harmonic gauge. The generalized $\tau^{ab}$ is not symmetric.

In \cite{BS}, Bi\v{c}\'ak and Schmidt (continuing earlier works \cite{Bicak,Bicak2} and inspired by \cite{BHL1,BHL2})
carried out a comprehensive search for energy-momentum tensors
under minimal conditions, and found multiparameter families
of energy-momentum tensors, depending on the gauge fixing conditions.
The gauges considered in \cite{BS}, besides no gauge fixing,
are the generalized harmonic gauges, as they have an important role in \cite{BHL1}
and are well suited to linearized gravity. 
Remarkably, in generalized harmonic gauges a larger family of energy-momentum tensors was found than without gauge fixing. 
Energy-momentum tensors are required to be homogeneous quadratic expressions in $\partial_c h_{ab}$ in \cite{BS},
but they are not required to be symmetric. The main method applied in \cite{BS}
is a direct method going back to the work of Fock \cite{Fock1,Fock2};
it is based on the field equations and avoids variational techniques and symmetry considerations,
therefore it is an additional task to connect the results it yields with symmetries. 
Some discussion of the connection of the obtained multiparameter families of tensors with spacetime translations is
included in \cite{BS}, nevertheless this can be made more complete.

In \cite{Barnett} (see also \cite{AAB}), a Maxwellian formulation of linearized gravity was
given by Barnett in transverse traceless (TT) gauge, and a positive energy density was found that differs from that of \cite{BHL1}.
This energy density can be taken to be a hint that there may exist other energy-momentum tensors than $\tau^{ab}$
that satisfy the d.e.c.\ in TT gauge.
Furthermore, the formulation of linearized gravity in \cite{Barnett} and the Fierz formalism specialized to TT gauge
are closely related, as will be discussed in subsequent sections.
It should be noted that for fields in TT gauge the energy densities of \cite{Barnett} and \cite{BHL1}
also appear as Hamiltonian densities in the earlier work \cite{DS2} of Deser and Seminara,
and the energy density of \cite{Barnett} also appears in the paper \cite{AfSt} of Afanasiev and Stepanovsky.

$T_{\mathrm{lg}}^{ab}$ will be a quadratic expression in $\partial_c h_{ab}$,
which is a common feature of the above-mentioned energy-momentum expressions in \cite{Barnett, BHL1, BS} as well;
this is one of the main reasons for the relevance of the results of \cite{Barnett, BHL1, BS} to our investigation.
$\tau^{ab}$ is especially interesting because of its favourable properties,
whereas the results of \cite{BS} provide a background and are useful for establishing uniqueness properties
of special energy-momentum tensors.
$T_{\mathrm{lg}}^{ab}$ turns out, as can be expected, to be a member of the general
multiparameter family of tensors obtained in \cite{BS} without gauge fixing.

Although in this paper we do not consider energy-momentum tensors that depend on higher than first derivatives of $h_{ab}$,
we note that
the results of \cite{BS} (for massless gravity without gauge fixing)
were extended in \cite{Baker} by Baker, also allowing terms of the type $h\partial\partial h$ in the energy-momentum tensor,
motivated by the fact that many expressions in the literature contain such terms
(see, for example, \cite{NN,BHLbootstrapping,BS,MagSok,Weinberg}).
In \cite{BS}, the divergence of the sought tensors was assumed to take a certain general form,
whereas in \cite{Baker} those tensors were determined
that can be obtained by correcting
the canonical energy-momentum tensor following from the Fierz--Pauli Lagrangian
by adding divergences of superpotentials and terms proportional to the equations of motion.

We do not require the energy-momentum tensor of the linearized gravitational field to be symmetric,
because this property is not necessary and it is not universally required in the literature.
A symmetric energy-momentum tensor is of interest in the context of the problem of giving the
energy-momentum tensor of the linearized gravitational field an equal footing with the energy-momentum tensor of matter
(see \cite{BHL2,Gupta,Kraichnan,DeserBootstrap,FeMoWa,BaGr,Padmanabhan,BHLbootstrapping,DeserBootstrap2,Weinberg}),
but this involves considering higher order approximations to the gravitational field equations
and is beyond the scope of the present paper.

Complete gauge invariance will not be required either.
Although it is a physically highly desirable property,
it is not crucial. Requiring it would be too rigorous;
no gauge invariant, local and Lorentz covariant
energy-momentum tensor of the linearized gravitational field is known to date,
and general results showing that a gauge invariant energy-momentum tensor of the linearized gravitational field
does not exist are available \cite{DeMcC,MagSok}.

Since $F_{abc}$ is antisymmetric in the first two indices, it is a natural question whether there exists a duality symmetry
of linearized gravity based on $F_{abc}$, rather than on the Riemann tensor (for the latter, see \cite{HT}).
We show that such a duality does exist,
and is closely related to the known duality based on the Riemann tensor.
It is defined only for fields in particular gauges; the most general gauge in which it can be defined is 
the $F_{ab}{}^b=0$ gauge, which is a special generalized harmonic gauge.
We also show that it can be restricted to traceless harmonic gauge, to TT gauge,
and to another gauge that is well suited to the Fierz formalism.
For fields in TT gauge it is close to the duality established in \cite{DS} (see also \cite{DS2})
by Deser and Seminara,
nevertheless our derivation of its existence as a transformation of $h_{ab}$
is quite different from the derivation presented in \cite{DS,DS2}.
A duality is also defined in TT gauge
in \cite{Barnett} by Barnett, and in this gauge it turns out to coincide with the duality introduced in this paper. 
We note that in \cite{Barnett} a helicity density associated with duality symmetry was constructed,
which was extended more recently into a conserved current in \cite{AAB},
using the same Maxwell-like formulation of linearized gravity as the one introduced in \cite{Barnett}.

In addition to applying the Fierz formalism,
we also review it in the framework of Lagrangian field theory.
This is done to explain in detail what the Fierz formalism is,
to show how one is led to it,
and because the only review of it known to us, presented in \cite{NN} (see also \cite{NN2}),
appears to need some correction.
The sections that contain our review are
Sections \ref{sec.lg.lagr}, \ref{sec.lg.foe} and Appendices \ref{app.dec}, \ref{app.riem}.
The discussion of the formalism is continued in Section \ref{sec.harmonic}
in generalized harmonic gauges and in related gauges, including the TT gauge.
The correction to \cite{NN} is discussed in Appendix \ref{app.nn}.

In Section \ref{sec.gauge}, we briefly recall Noether's theorem and apply it to determine
the conserved currents associated with the gauge symmetry of linearized gravity.
These currents are used in Section \ref{sec.emtensor}.

In Section \ref{sec.emtensor}, the energy-momentum tensor of the linearized gravitational field
without gauge fixing is discussed.
In particular, $T_{\mathrm{lg}}^{ab}$ is introduced.

In Section \ref{sec.harmonic}, the discussion of the field equations and the energy-momentum tensor
is continued in generalized harmonic gauges and in other related gauges.

In Section \ref{sec.emtsum}, a summary of the main results concerning $T_{\mathrm{lg}}^{ab}$ is given.  

In Section \ref{sec.lg.duality}, the duality symmetry of the linearized Einstein equations
under which the Fierz tensor is mapped into its dual is introduced.
A conserved current associated with duality symmetry is constructed,
and the action of duality on monochromatic plane waves is determined.

In Section \ref{sec.concl}, concluding remarks on the results are made.\\

\noindent
Spacetime tensor indices, which can take the values $0,1,2,3$, are denoted by $a,b,\dots$,
while spatial tensor indices, taking the values $1,2,3$, are denoted by $i,j,\dots$.
The Minkowski metric, with signature $(+,-,-,-)$, will be denoted by $\eta_{ab}$.
Spacetime tensor indices are raised and lowered by $\eta_{ab}$.
Spatial tensor indices are raised and lowered by the spatial Euclidean metric $\eta_{ij}$, which has the signature $(-,-,-)$.
Tensor components are understood to be taken with respect to normalized orthogonal basis vectors.
$\epsilon^{abcd}$ will denote the rank four totally antisymmetric tensor, normalized so that
$\epsilon^{0123} = -\epsilon_{0123} = 1$. 
The spatial rank three totally antisymmetric tensor will be denoted by $\epsilon^{ijk}$, and
$\epsilon^{123} = -\epsilon_{123} = 1$. 
Contracted pairs of indices of a single tensor will often be omitted (e.g., $h$ will be written instead of $h_a{}^a$).
A tilde will be used to denote the Hodge dual of a tensor $T^{ab\dots}$ that is antisymmetric in the first two indices: 
$\tilde{T}^{ab\dots}=\frac{1}{2}\epsilon^{abcd}T_{cd}{}^{\dots}$.
As is well known, $\tilde{\tilde{T}}^{ab\dots}=-T^{ab\dots}$.
$\Box$ will denote the wave operator $\partial_a\partial^a$.
The $k$-th de Rham cohomology group of a manifold $M$ will be denoted by $H^k_{\mathrm{dR}}(M)$.
We shall use natural units in which the speed of light is $1$ and $8\pi\gamma=1$,
where $\gamma$ is Newton's gravitational constant.

Let $\Phi_\alpha$ be a collection of fields, labeled by an index $\alpha$.
A function $f$ on the spacetime will be said to be a local function of $\Phi_\alpha$
if it is of the composite form
$f(\Phi_\alpha(x^a),\partial_{a_1}\Phi_\alpha(x^a),\\ \partial_{a_1a_2}\Phi_\alpha(x^a),\dots,\partial_{a_1a_2\dots a_k}\Phi_\alpha(x^a),x^a)$
with some number $k$.
The notation $f[\Phi_\alpha]$ will also be used for local functions.
$\partial_a f$ will be understood to be the partial derivative of $f$ as a composite function.
The Euler--Lagrange derivative $\frac{\partial L}{\partial\Phi_\alpha}-\partial_a\frac{\partial L}{\partial(\partial_a\Phi_\alpha)}$
of a Lagrangian density $L(\Phi_\alpha(x^a),\partial_a\Phi_\alpha(x^a),x^a)$ will be denoted by $\mr{E}(L)^\alpha$.

\section{Formulation of linearized gravity using the Fierz tensor}
\label{sec.lg.lagr}

The linearized gravitational field $h_{ab}$ is defined by $g_{ab}=\eta_{ab}+h_{ab}$, 
where $g_{ab}$ is the complete metric, $\eta_{ab}$ is the Minkowski metric, 
and $h_{ab}$ is assumed to be small so that the linearized Einstein field equations
\beq
\label{eq.linein}
G^{ab}=\mc{T}^{ab}
\eeq
are a good approximation. In (\ref{eq.linein}), $G^{ab}$ is the linearized Einstein tensor
and $\mc{T}^{ab}$ is the
matter energy-momentum tensor of the linearized theory. 
$\mc{T}^{ab}$ is assumed to be a divergenceless symmetric tensor field.
For explicit expressions for $G_{ab}$ and for the 
linearized Riemann tensor, Ricci tensor, Ricci scalar and Christoffel symbols, which will be 
denoted by $R_{abcd}$, $R_{ab}$, $R$ and $\Gamma_{abc}$, see Appendix \ref{app.riem}.
$h_a{}^a$ will be denoted by $h$.
$h_{ab}$ will be assumed to be defined either on the entire Minkowski spacetime or on some part of it.
In this paper we shall study the theory defined by (\ref{eq.linein}) in itself,
without attempting to find counterparts of the various quantities
in full nonlinear general relativity.

The usual Lagrangian density function
for the linearized gravitational field without source is the Fierz--Pauli Lagrangian with zero mass:
\bea
L_{\mathrm{FP}} & = &
\frac{1}{4}\partial_a h_{bc}\partial^a h^{bc} 
-\frac{1}{2}\partial_c h^{ba}\partial_b h^c{}_a
+\frac{1}{2}\partial_a h^{ab} \partial_b h
-\frac{1}{4}\partial_a h \partial^a h \\
& = & \Gamma^c{}_a{}^a  \Gamma^b{}_{cb} 
- {\Gamma^c}_{ab}  \Gamma^b{}^a{}_c.
\label{eq.lg.fp}
\eea
$L_{\mathrm{FP}}$ is equivalent with the Lagrangian $\frac{1}{2}h^{ab}G_{ab}$
in the sense that they differ by a total divergence.
The Euler--Lagrange equations following from $L_{\mathrm{FP}}$ are the linearized Einstein equations
in vacuum:
\beq
\mr{E}(L_{\mathrm{FP}})^{ab} =G^{ab}= R^{ab}-\frac{1}{2}\eta^{ab}R =
\partial_c[\Gamma^{cab} -\eta^{bc}\Gamma^{da}{}_d+\frac{1}{2}\eta^{ab}\Gamma^{dc}{}_d
-\frac{1}{2}\eta^{ab}\Gamma^c{}_d{}^d] = 0.
\label{eq.lg.le}
\eeq
$\Gamma_{abc}$ is often regarded as the gravitational field strength, and then
$L_{\mathrm{FP}}$ in the form (\ref{eq.lg.fp}) and the linearized Einstein equations bear some resemblance to the
usual Lagrangian $-\frac{1}{4}F_{ab}F^{ab}$ of the electromagnetic field and to the inhomogeneous Maxwell equations.
($F_{ab}=\partial_a A_b - \partial_b A_a$, where $A_a$ denotes the vector potential.)

In order to find a Lagrangian that is more similar to the standard Lagrangian of electrodynamics than $L_{\mathrm{FP}}$, 
let us first determine the Lagrangians that are Lorentz covariant, 
differ from $L_{\mathrm{FP}}$ only by a total divergence and are
homogeneous quadratic expressions in $\partial_a h_{bc}$ with constant coefficients.
One finds that the Lagrangians with these properties constitute a $2$-parameter family,
spanned by $L_{\mathrm{FP}}$ and
\bea
L_2 & = & \partial_c h^{ba}\partial_b h^c{}_a - \partial_b h^{ba}\partial_c h^c{}_a
 \ = \ \partial_c (h^{ba}\partial_b h^c{}_a - h^{ca}\partial_b h^b{}_a),
\label{eq.lg.l2} \\
L_3 & = & \frac{1}{4}\epsilon^{abcd}\partial_a h_{be} \partial_c h_d{}^e \ = \ 
\frac{1}{4}\partial_a (\epsilon^{abcd}h_{be} \partial_c h_d{}^e ).
\label{eq.lg.l3}
\eea
For a reason explained below, it is useful to introduce the Lagrangian
\beq
L_1 = L_{\mathrm{FP}}+\frac{1}{4}L_2 =   \frac{1}{4}\partial_a h_{bc}\partial^a h^{bc} 
-\frac{1}{4}(\partial_c h^{ba}\partial_b h^c{}_a + \partial_b h^{ba}\partial_c h^c{}_a)
+\frac{1}{2}\partial_a h^{ab} \partial_b h
-\frac{1}{4}\partial_a h \partial^a h.
\label{eq.lg.l1}
\eeq
The general $2$-parameter Lagrangian with the properties above can then be written as
\beq
L=L_1+\alpha_2 L_2+\alpha_3 L_3,
\label{eq.lg.l}
\eeq
where $\alpha_2$ and $\alpha_3$ are the free parameters. 
Since $L_2$ and $L_3$ are total divergences,\\
$\mr{E}(L)^{ab} = G^{ab}$, i.e.\ $L$ generates the same field equations
for any $\alpha_2$, $\alpha_3$.

Next, we consider the gauge transformation properties of $L$. 
$G^{ab}$ is invariant under the gauge transformations
\beq
\label{eq.lg.g}
h_{ab} \to h_{ab}+ \partial_a\phi_b+\partial_b\phi_a,
\eeq
where $\phi_a$ is an arbitrary covector field. This symmetry is known to be a remnant of the diffeomorphism 
symmetry of full general relativity. However, Poincar\'e transformations are not members of the symmetry group
constituted by these gauge transformations.
In contrast with the case of electrodynamics, $L$ is invariant under (\ref{eq.lg.g}) only 
up to a total divergence for any values of $\alpha_2$ and $\alpha_3$.
Nevertheless, it is possible to find Lagrangians that are strictly invariant 
(i.e., not only up to total divergence)
under the special gauge transformations in which the parameter
$\phi_a$ takes the form $\phi_a=\partial_a\phi$, where $\phi$ is an arbitrary scalar function.
Specifically, $L_1$ and $L_3$ are strictly invariant under these gauge transformations,
which will be called \emph{scalar gauge transformations} hereafter\footnote{It would also be possible to define scalar gauge transformations
as gauge transformations whose parameter satisfies the condition $\partial_a\phi_b-\partial_b\phi_a=0$.}.
We have introduced $L_1$ because of this invariance property;
$L_{\mathrm{FP}}$ is not strictly invariant under scalar gauge transformations.

A further property of the standard Maxwell Lagrangian is that it is expressed 
in terms of a field strength tensor, namely $F_{ab}$,
which is gauge invariant and is a linear expression in $\partial_a A_b$,
i.e.\ in the first derivative of the basic dynamical field. We should therefore find a suitable counterpart
of $F_{ab}$ in linearized gravity.
In linearized gravity a gauge invariant field strength tensor
that does not depend on higher than first derivatives of $h_{ab}$ does not exist,
nevertheless the situation is better if only invariance under scalar gauge transformations is demanded.
Let us consider the tensor
\beq
F_{abc} = \frac{1}{2} (\partial_a h_{bc}-\partial_b h_{ac}  
+\partial_d h^d{}_a\eta_{bc}-\partial_d h^d{}_b\eta_{ac}
-\partial_a h\hspace{0.3mm}\eta_{bc}+\partial_b h\hspace{0.3mm}\eta_{ac} ),
\label{eq.lg.x6}
\eeq
called \emph{Fierz tensor} in \cite{NN}.
$F_{abc}$ is invariant under scalar gauge transformations,
it is obviously antisymmetric in the first two indices,
\beq
F_{abc}=-F_{bac},
\label{eq.lg.x1}
\eeq
and has the cyclic property
\beq
F_{abc}+F_{bca}+F_{cab}=0,
\label{eq.lg.x2}
\eeq
which, taking into account (\ref{eq.lg.x1}), can also be written as
\beq
-\frac{1}{2}\epsilon^{abcd}F_{abc}=\tilde{F}^d{}_a{}^a = 0.
\label{eq.lg.x2b}
\eeq
Furthermore, it is possible to express $L_1$ in terms of $F_{abc}$:
\beq
L_1=\frac{1}{2}( F_{abc}F^{abc} - F_a F^a ),\qquad F_a \equiv F_{ab}{}^b = \partial_b h^b{}_a - \partial_a h,
\label{eq.lg.x7}
\eeq
and the form of this expression for $L_1$ shows remarkable similarity to the standard
Maxwell Lagrangian.
Nevertheless, a new feature of $F_{abc}$ in comparison with the electromagnetic field strength is that
$F_{abc}$ has trace in the last two indices,
which enters $L_1$.

We note that 
\beq
\label{eq.lg.xx1}
F_{abc}F^{acb}= \frac{1}{2}F_{abc}F^{abc}
\eeq
\beq
\label{eq.lg.xx2}
F_{ab}{}^c F^{dab}= F_{ab}{}^d F^{cab} = -\frac{1}{2}F_{ab}{}^c F^{abd},
\eeq
which follow from (\ref{eq.lg.x1}) and (\ref{eq.lg.x2}).
(\ref{eq.lg.xx1}) shows that there are two independent quadratic scalars that can be constructed from $F_{abc}$ without using 
$\epsilon^{abcd}$, namely $F_{abc}F^{abc}$ and $F_a F^a$.

$F_{abc}$ also satisfies the identities
\beq
\partial_c F^c{}_{ab}=-G_{ab},\qquad
\partial_c F_{ab}{}^c=0.
\label{eq.lg.x3}
\eeq
In particular,
\beq
\partial_c F^{cab}= \partial_c F^{cba}.
\eeq

From (\ref{eq.lg.x1}) it follows that
\beq
\label{eq.lg.dualcycl}
\epsilon^{dabc}\tilde{F}_{abc} = -2F^d.
\eeq
This implies that $\tilde{F}_{abc}$ has the cyclic property (\ref{eq.lg.x2}) only if $F_a=0$.

Regarding the Euler--Lagrange equations, we have 
\beq
\frac{\partial L_1}{\partial (\partial_c h_{ab})}=\frac{1}{2}(F^{cab}+F^{cba}),
\label{eq.lg.dl1ddg}
\eeq
thus the Euler--Lagrange equations take the form
\beq
\label{eq.lg.x8}
\mr{E}(L_1)^{ab} =
-\partial_c F^{cab}=0.
\eeq
In the presence of matter, the last equation becomes
\beq
\label{eq.lg.x9}
-\partial_c F^{cab} = \mc{T}^{ab},
\eeq
taking into account (\ref{eq.linein}) and (\ref{eq.lg.x3}).
This equation is clearly similar to the inhomogeneous Maxwell equation.

Summarizing the above developments, it can be said that
by introducing $L_1$, $F_{abc}$ and the concept of scalar gauge transformations,
the Lagrangian formulation of linearized gravity
can be brought to a form that shows good analogy with electrodynamics in its usual relativistic form,
although significant differences between the two theories remain. 
$F_{abc}$ appears to be a better analogue of the electromagnetic field strength than $\Gamma_{abc}$, 
because it is antisymmetric in the first two indices,
it is invariant under scalar gauge transformations (in contrast with $\Gamma_{abc}$), 
there exists a Lagrangian expressed in terms of $F_{abc}$ (i.e., $L_1$)
that has strong resemblance to the usual Lagrangian of electrodynamics,
and the linearized Einstein equations also take a form in terms of $F_{abc}$ that is 
very similar to the inhomogeneous Maxwell equations.

In the next subsection, in Section \ref{sec.harmonic}
and in Appendices \ref{app.dec}, \ref{app.riem},
we further discuss the above formulation of linearized gravity and add some extensions to it.

\subsection{Further details of the formalism}
\label{sec.dev}

Besides $F_{abc}$, it is useful to introduce the tensor
\beq
\label{eq.circx}
\mathring{F}_{abc}=
\frac{1}{2}(\partial_a h_{bc}-\partial_b h_{ac})
\eeq
as well.
$\mathring{F}_{abc}$ and $F_{abc}$ are equivalent in the sense that
they can be expressed algebraically in terms of each other 
(see Appendix \ref{app.dec}) and
$\mathring{F}_{abc}$ also has the properties (\ref{eq.lg.x1}), (\ref{eq.lg.x2}). 
As a consequence of this relationship, $\mathring{F}_{abc}$ is also invariant under scalar gauge transformations.
Using $\mathring{F}_{abc}$, $L_1$ can be written in the following additional forms:
\beq
\label{eq.L1forms}
L_1 = \frac{1}{2}\mathring{F}_{abc}\mathring{F}^{abc} - \mathring{F}_a\mathring{F}^a
= \frac{1}{2}F_{abc}\mathring{F}^{abc}.
\eeq
Moreover, from (\ref{eq.circx}) it clearly follows that $\mathring{F}_{abc}$ satisfies the identity
\beq
\label{eq.circ3}
\partial_c \tilde{\mathring{F}}^{cab}=0,
\eeq
which can be regarded as an analogue of the homogeneous Maxwell equation $\partial_a \tilde{F}^{ab}=0$.
This analogy will be discussed in more detail in Section \ref{sec.lg.foe}.  
The expression $\frac{1}{2}F_{abc}\mathring{F}^{abc}$ for $L_1$ harmonizes remarkably well with
the equations (\ref{eq.lg.x8}), (\ref{eq.circ3})
(see also (\ref{eq.lg.foe})).
To our knowledge, this form of $L_1$ has not previously appeared in the literature.

A further new feature of $F_{abc}$ in comparison with the electromagnetic field tensor, 
related to the fact that it has a trace, is that 
it can be decomposed into traceless and trace parts, which we denote by $\breve{F}_{abc}$ and $\check{F}_{abc}$.
The details of this decomposition are discussed in Appendix \ref{app.dec}.

As is well known, $\partial_c h_{ab}$ can be expressed in terms of $\Gamma_{abc}$. In contrast with this,
$\partial_c h_{ab}$ (and thus also $\Gamma_{abc}$) cannot be expressed in terms of $F_{abc}$, since $F_{abc}$
is invariant under scalar gauge transformations.
The analogous fact in electrodynamics is that $\partial_a A_b$ is not gauge invariant
and cannot be expressed in terms of $F_{ab}$. Since $\Gamma_{abc}$ cannot be expressed in terms of
$F_{abc}$, it is a nontrivial fact that there exists a Lagrangian (namely $L_1$)
that can be expressed in terms of $F_{abc}$.

The linearized Riemann tensor
is completely gauge invariant,
therefore it is natural to ask if, along with the linearized Ricci tensor and Ricci scalar,
it can be expressed in terms of $F_{abc}$.
As one expects, this is indeed possible; see Appendix \ref{app.riem} for the details.

The linearized Weyl tensor can be expressed in terms of $\breve{F}_{abc}$;
see (\ref{eq.weyl2}), (\ref{eq.weyl3}).
In view of (\ref{eq.weyl3}) and the algebraic properties $\breve{F}_{abc}$ has,
$\breve{F}_{abc}$ can be regarded as a Lanczos potential for the linearized Weyl tensor \cite{Lanczos}-\cite{JWK}.
We stress, however, that in the present setting
$\breve{F}_{abc}$
is primarily a field strength, rather than a potential.

In \cite{JWK} it is pointed out that Lanczos potentials,
and generally tensors having the algebraic properties (\ref{eq.lg.x1}), (\ref{eq.lg.x2}),
have an equivalent symmetric form. For $F_{abc}$, this form can be defined as 
$\cF_{cab} = F_{cab} + F_{cba}$.
$\cF_{cab}$ has the algebraic properties
$\cF_{cab}=\cF_{cba}$,
$\cF_{cab}+\cF_{abc}+\cF_{bca} = 0$.
The inverse relation between $F_{abc}$ and $\cF_{cab}$ is
$F_{abc}=\frac{1}{3}(\cF_{abc}-\cF_{bac})$.

$L_3$ can be written as 
\beq
\label{eq.lg.l3b}
L_3 = \frac{1}{4}\epsilon^{abcd}\breve{F}_{abe}\breve{F}_{cd}{}^e.
\eeq
There are also several other similar forms of $L_3$ in virtue of (\ref{eq.xort3}), for example\\
$L_3 = \frac{1}{4}\epsilon^{abcd}\mathring{F}_{abe}\mathring{F}_{cd}{}^e$.

The second identity in (\ref{eq.lg.x3}) shows that $F_{abc}$ can be expressed as a divergence of a rank four tensor:
\beq 
\label{eq.lg.H1}
F_{abc}=\partial_d H_{abc}{}^d.
\eeq
In particular,
\beq
\label{eq.lg.H2}
H_{abcd}=\frac{1}{2} (h_{bc}\eta_{ad} + h_{ad}\eta_{bc} - h_{ac}\eta_{bd} - h_{bd}\eta_{ac}
+ h (\eta_{ac} \eta_{bd} - \eta_{bc} \eta_{ad}))
\eeq
is suitable.
This $H_{abcd}$ has the algebraic symmetry properties
\beq
\label{eq.lg.H3}
H_{abcd}=-H_{bacd},\quad H_{abcd}=-H_{abdc},\quad H_{abcd}=H_{cdab}, 
\eeq
\beq
\label{eq.lg.H4}
H_{abcd}+H_{bcad}+H_{cabd}=0,
\eeq
and appears in several articles; see, for example, \cite{AD, Jezierski, BB, AS}.

In electrodynamics it is well known that if $F_{ab}=0$ on a spacetime domain that has zero first de Rham cohomology,
then $A_a$ is zero on that domain up to a gauge transformation.
For the Fierz tensor the following analogous statement can be made:
\begin{lemma}
\label{Fabc0}
If $F_{abc}=0$ on a spacetime domain $\Omega$ and $H^1_{\mathrm{dR}}(\Omega)=0$,
then $h_{ab}=\partial_{ab}\phi$ on $\Omega$ with some function $\phi$
(i.e., $h_{ab}$ is zero up to scalar gauge transformation).
\end{lemma}
To prove this statement, we first note that $F_{abc}=0$ implies $F_a=\mathring{F}_a=0$ and thus
$\mathring{F}_{abc}=0$ (see Appendix \ref{app.dec}). In view of (\ref{eq.circx}),
$\mathring{F}_{abc}=0$ implies, together with $H^1_{\mathrm{dR}}(\Omega)=0$,
that there exists a covector field $\phi_a$ so that $h_{ab}=\partial_a\phi_b$ on $\Omega$.
Since $h_{ab}$ is symmetric,
$\partial_a\phi_b-\partial_b\phi_a=0$. From the latter equation it follows,
again taking into account $H^1_{\mathrm{dR}}(\Omega)=0$,
that there exists a function $\phi$ so that $\phi_a=\partial_a\phi$ on $\Omega$,
and thus $h_{ab}=\partial_{ab}\phi$ on $\Omega$.
\hfill $\blacksquare$

\section{First order Maxwell-like field equations}
\label{sec.lg.foe}

It is well known that from the homogeneous Maxwell equations $\partial_a \tilde{F}^{ab}=0$,
regarded as equations for $F_{ab}$, 
it follows that $F_{ab}$ has a vector potential 
in any spacetime domain $\Omega$ that has zero second de Rham cohomology.
This implies that in such spacetime domains
the field equations $\Box A^b -\partial^b\partial_a A^a = \mc{J}^b$ of electrodynamics
(where $\mc{J}^b$ denotes the electric current)
are equivalent with the set of first order differential equations 
$\partial_a F^{ab}=\mc{J}^b$, $\partial_a \tilde{F}^{ab}=0$, 
where $F_{ab}$ is understood to be the basic field variable instead of the vector potential. 
The aim of this section is to show that the linearized Einstein equations admit a similar first order formulation.
The basic field variable in this formulation is $F_{abc}$ with the properties (\ref{eq.lg.x1}), (\ref{eq.lg.x2}),
and $\mathring{F}_{abc}$
is understood to be defined as
$\mathring{F}_{abc} = F_{abc} - \frac{1}{2} (F_a\eta_{bc}-F_b\eta_{ac})$,
in agreement with (\ref{eq.circ2}) and (\ref{eq.trx1}).
The inverse relation between $F_{abc}$ and $\mathring{F}_{abc}$ is given by (\ref{eq.circ4}).

The main result that allows one to write the field equations of linearized gravity
in Maxwell-like first order form is the following:
\begin{lemma}
\label{pr1}
Let $E_{abc}$ be a tensor field with the algebraic symmetries
$E_{abc} = - E_{bac}$,\\
$E_{abc}+E_{bca}+E_{cab}=0$, defined on a spacetime domain $\Omega$.
If $\partial_c \tilde{E}^{cab}=0$ and $H^2_{\mathrm{dR}}(\Omega)=0$,
then there exists a symmetric tensor field $h_{ab}$, defined on $\Omega$,
so that $E_{abc}=\frac{1}{2}(\partial_a h_{bc} - \partial_b h_{ac})$.
\end{lemma}
Lemma \ref{pr1} can be proved in the following way: $\partial_c \tilde{E}^{cab}=0$
together with $H^2_{\mathrm{dR}}(\Omega)=0$ implies that
$E_{abc}=\frac{1}{2}(\partial_a C_{bc}-\partial_b C_{ac})$
with some tensor field $C_{ab}$, which is not necessarily symmetric.
Let $A_{ab}$ and $B_{ab}$ be the antisymmetric and symmetric part, respectively, of $C_{ab}$, so that $C_{ab}=A_{ab}+B_{ab}$.
Then $E_{abc}+E_{bca}+E_{cab}= \partial_a A_{bc} + \partial_b A_{ca} + \partial_c A_{ab}$.
This has to be zero due to the algebraic properties of $E_{abc}$,
implying, in view of $H^2_{\mathrm{dR}}(\Omega)=0$,
that $A_{ab}$ has a vector potential, i.e.\ $A_{ab}=\partial_a H_b - \partial_b H_a$ with 
some $H_a$. Using this vector potential, $E_{abc}$ can be written as
$E_{abc}= \frac{1}{2}(\partial_a B_{bc}-\partial_b B_{ac} + \partial_{bc} H_a - \partial_{ac}H_b)
= \frac{1}{2}\bigl(\partial_a (B_{bc}- \partial_b H_c - \partial_c H_b) 
- \partial_b (B_{ac}- \partial_a H_c - \partial_c H_a)\bigr)$, therefore 
$B_{ab}- \partial_a H_b - \partial_b H_a$ is a suitable choice for $h_{ab}$.
\hfill$\blacksquare$\\

The relation between $F_{abc}$ and $\mathring{F}_{abc}$ given by (\ref{eq.circ2}) and (\ref{eq.trx1})
implies that if $F_{abc}$ has the form (\ref{eq.lg.x6}), then $\mathring{F}_{abc}$ has the form (\ref{eq.circx}),
and therefore $\partial_c\tilde{\mathring{F}}^{cab} = 0$ (see (\ref{eq.circ3})).
From Lemma \ref{pr1} it follows that the converse is also true if $H^2_{\mathrm{dR}}(\Omega)=0$:
$\partial_c\tilde{\mathring{F}}^{cab} = 0$ implies that (\ref{eq.circx}) holds with some symmetric tensor field $h_{ab}$,
and then the relation (\ref{eq.circ4}) between $F_{abc}$ and $\mathring{F}_{abc}$ implies that (\ref{eq.lg.x6}) holds for $F_{abc}$.
This means that a possible form of the field equations of the linearized gravitational field is
\beq 
\label{eq.lg.foe}
-\partial_c F^{cab} = \mc{T}^{ab},\qquad\quad \partial_c\tilde{\mathring{F}}^{cab} = 0,
\eeq
where the basic field variable is understood to be $F_{abc}$ (which is assumed to have the properties 
(\ref{eq.lg.x1}) and (\ref{eq.lg.x2})), and $\mathring{F}_{abc}$ is understood to be
defined as\\
$\mathring{F}_{abc} = F_{abc} - \frac{1}{2} (F_a\eta_{bc}-F_b\eta_{ac})$.
These equations are equivalent with the linearized Einstein equations in spacetime domains
that have zero second de Rham cohomology. 
They are clearly analogous to Maxwell's equations in their first order form,
although they are less symmetric,
as the homogeneous equations contain $\mathring{F}_{abc}$ instead of $F_{abc}$. 
They become symmetric in suitable gauges, nevertheless---see Section \ref{sec.harmonic}.\\

The identity $\partial_c F^{abc}=0$ (see (\ref{eq.lg.x3})) can be derived from $\partial_c\tilde{\mathring{F}}^{cab}=0$
directly (i.e., without using Lemma \ref{pr1}):
from $\partial_a \mathring{F}_{bcd} + \partial_b \mathring{F}_{cad} + \partial_c \mathring{F}_{abd} = 0$
one gets $\partial_c \mathring{F}_{ab}{}^c + \partial_a \mathring{F}_b - \partial_b \mathring{F}_a = 0$ by contracting
$c$ with $d$, and then by using (\ref{eq.circ4}) one obtains $\partial_c F_{ab}{}^c = 0$.

$\partial_c F^{cab}=\partial_c F^{cba}$ can also be derived from $\partial_c\tilde{\mathring{F}}^{cab}=0$
directly: from the property (\ref{eq.lg.x2}) of $F^{abc}$ it follows that
$\partial_c F^{abc} + \partial_c F^{bca} + \partial_c F^{cab} = 0$,
which implies $\partial_c F^{bca} + \partial_c F^{cab} = 0$, taking into account $\partial_c F_{ab}{}^c = 0$,
and then one gets $\partial_c F^{cab}=\partial_c F^{cba}$ using (\ref{eq.lg.x1}).

In \cite{NN} it is claimed that $\partial_a\tilde{F}^{abc}+\partial_a\tilde{F}^{acb}=0$ is the condition for the existence
of a symmetric tensor potential $h_{ab}$ for $F^{abc}$. This is examined in detail in Appendix \ref{app.nn}.

\subsection{Wave equation for the field strength tensor}
\label{sec.lg.wave}

It is well known that from the Maxwell equations
$\partial_a F^{ab}=\mc{J}^b$, $\partial_a \tilde{F}^{ab}=0$ it follows that $F_{ab}$ satisfies the inhomogeneous
wave equation $\Box F_{ab}=\partial_a\mc{J}_b - \partial_b \mc{J}_a$.
A similar result cannot be expected for $F_{abc}$ or $\mathring{F}_{abc}$, 
because they are not fully gauge invariant.
Nevertheless, from (\ref{eq.lg.foe}) it follows that 
\beq
\label{eq.lg.fwav}
\Box \mathring{F}_{abc} = - (\partial_a \mc{T}_{bc} - \partial_b\mc{T}_{ac})
-\frac{3}{2} (\partial_{ad} \check{F}^d{}_{bc}-\partial_{bd}\check{F}^d{}_{ac}),
\eeq
where the source term depends on $F_{abc}$ only through the trace part $\check{F}_{abc}$.
\newpage
In order to derive 
(\ref{eq.lg.fwav}), one takes the divergence of the second equation in (\ref{eq.lg.foe}):
$\partial^d(\partial_d\mathring{F}_{abc}+ \partial_a\mathring{F}_{bdc} +\partial_b \mathring{F}_{dac})=0$,
which can be written as
$\Box \mathring{F}_{abc} - \partial_{ad}\mathring{F}^d{}_{bc} +\partial_{bd} \mathring{F}^d{}_{ac}=0$,
and then by using (\ref{eq.circ2}) and the first equation in (\ref{eq.lg.foe}) one obtains (\ref{eq.lg.fwav}).

By taking the trace of (\ref{eq.lg.fwav}), one gets the equation
$\partial_{ab}F^b = - \partial_a \mc{T}$ for the trace of $F_{abc}$,
which is just the derivative of the trace
\beq
\label{eq.eintr}
\partial_a F^a = - \mc{T}
\eeq
of the linearized Einstein equations $\partial_c F^{cab} = -\mc{T}^{ab}$.

Although (\ref{eq.lg.fwav}) is not an ideal wave equation because of the presence of
$-\frac{3}{2}(\partial_{ad} \check{F}^d{}_{bc}-\partial_{bd}\check{F}^d{}_{ac})$ on the right hand side,
it is possible to find a proper wave equation for $F_{abc}$
in certain gauges, including the generalized harmonic gauges---see Section \ref{sec.harmonic} for the details.

\section{Conserved currents in linearized gravity associated with gauge symmetry}
\label{sec.gauge}

In this section the conserved currents corresponding to the gauge symmetry of linearized gravity are determined
applying Noether's theorem, and they are found to exhibit good analogy with
the comparable currents of electrodynamics.
Both electric and magnetic type currents are obtained. 
$\mc{T}^{ab}$, as conserved tensor field, is found to correspond to spacetime translation-like gauge transformations.

As Noether's theorem is important for this section as well as for the subsequent ones,
we start by very briefly recalling it, together with basic definitions concerning conserved currents.
In the literature several variants of Noether's theorem can be found;
in this paper we shall use, instead of the original variant, a newer one,
which is sufficiently general for our purpose and is remarkably simple.
Transformations are taken to act on the fields in this variant, leaving the coordinates unchanged.
In this way spacetime and internal symmetries are dealt with on the same footing. 
For references on Noether's theorem, see \cite{BR}-\cite{K-S}, among others;
for the variant we use, see \cite{BR,Pons}, for example.

Let $\Phi_\alpha$, where $\alpha$ is a general index,
be the collection of the basic dynamical fields in the theory that one intends to consider.
A \emph{strongly conserved current} is a current that is conserved
for arbitrary field configurations,
a \emph{weakly zero current} is a current that is zero
if the fields satisfy the field equations,
and a \emph{trivially conserved current} is a linear combination of
a strongly conserved current and a weakly zero current.
Two currents, $J_1^a$ and $J_2^a$, are said to be \emph{equivalent} (see \cite{Olver}) if
$J_1^a-J_2^a$ is a trivially conserved current.
A \emph{superpotential} for a strongly
conserved current $J^a$ is an antisymmetric tensor field $\Sigma^{ab}$ for which
$J^a=\partial_b \Sigma^{ab}$ holds for arbitrary field configurations.
We assume, when we do not state otherwise,
that currents and superpotentials are
local functions of $\Phi_\alpha$ (see the last paragraph of Section \ref{sec.intr}),
and we only consider strongly conserved currents that follow from a superpotential.
In virtue of Stokes' theorem, the charge integral $\int_D d^3x\, J^0$ over a spatial domain $D$ can be converted
to an integral over the boundary surface $\partial D$ if $J^a$ follows from a superpotential,
and the integrand will be a local function of $\Phi_\alpha$ (since the superpotential is assumed to be a local function of $\Phi_\alpha$).
If the integrand in the surface integral falls off sufficiently rapidly at infinity, then the \emph{total charge}
$\int_{\RR^3} d^3x\, J^0$ is zero. As a consequence, the total charges determined by two equivalent currents are equal
for field configurations satisfying suitable fall off conditions and the field equations.
This equality is one of the main justifications for the above definition of equivalence of currents.
The aforementioned definitions and statements can be extended in an obvious way to currents that have additional indices of any type.

Turning to Noether's theorem,
let $L(\Phi_\alpha,\partial_a\Phi_\alpha,x^a)$ be a Lagrangian density function for $\Phi_\alpha$.
Under an infinitesimal transformation $\Phi_\alpha\to \Phi_\alpha+\epsilon\delta\Phi_\alpha$ of $\Phi_\alpha$,
where $\epsilon$ is a small parameter,
the variation of $L$ is 
\beq
\label{eq.nthr1}
\delta L = \frac{dL[\Phi_\alpha + \epsilon \delta \Phi_\alpha]}{d\epsilon}|_{\epsilon = 0} 
= \frac{\partial L}{\partial \Phi_\alpha}\delta\Phi_\alpha 
+ \frac{\partial L}{\partial (\partial_a \Phi_\alpha)}\partial_a \delta \Phi_\alpha
= \mr{E}(L)^\alpha \delta\Phi_\alpha + \partial_a j^a,
\eeq
where $j^a = \frac{\partial L}{\partial (\partial_a \Phi_\alpha)}\delta \Phi_\alpha$.
If $\delta \Phi_\alpha$ is a local function of the basic fields
and $\delta L = \partial_a K^a$
holds for arbitrary configurations of $\Phi_\alpha$
with some $K^a$ that is also a local function of $\Phi_\alpha$, 
then the infinitesimal transformation characterized by the variation $\delta\Phi_\alpha$
is said to be a symmetry of $L$\footnote{The term \emph{Noether symmetry} is also used in the literature.},
and the current
$J^a = j^a-K^a$,
called \emph{Noether current}, is conserved (i.e., $\partial_a J^a=0$) for all field configurations that satisfy
the Euler--Lagrange equations $\mr{E}(L)^\alpha = 0$.
For arbitrary field configurations the identity
\beq
\label{eq.nthr2}
\partial_a J^a=- \mr{E}(L)^\alpha \delta\Phi_\alpha
\eeq
holds.
The sum of two Noether currents $J_1^a$, $J_2^a$, associated with $\delta_1\Phi_\alpha$, $\delta_2\Phi_\alpha$,
is a Noether current associated with $\delta_1\Phi_\alpha+\delta_2\Phi_\alpha$.
The $K^a$ quantity relevant for $J_1^a+J_2^a$ is also the sum of the $K^a$-s for $J_1^a$ and $J_2^a$.

The identity $\delta L=\partial_a K^a$ does not fix $K^a$ uniquely;
an arbitrary strongly conserved current can be freely added to it, and thus also to $J^a$.
Nevertheless it is often possible
to reduce this indeterminacy by imposing reasonable conditions on $K^a$.
For instance, if $L$ is strictly invariant, i.e.\ $\delta L = 0$, then $K^a$ is usually chosen to be zero,
and the ambiguity of $J^a$ is thereby completely eliminated.
It is also often reasonable to require that $K^a$ should not depend on higher than first derivatives of $\Phi_\alpha$.

By adding a weakly zero current to a Noether current we do not necessarily get another Noether current,
since a weakly zero current is not necessarily a Noether current.
Thus weakly zero currents generally cannot be added freely to a Noether current.

The Noether currents associated with gauge symmetries are trivially conserved currents.
Using the freedom in $K^a$, they can be chosen to be weakly zero,
but these weakly zero currents may depend on higher derivatives of the fields than some other choices.

In the case of spacetime translations,
if the variation of $\Phi_\alpha$ is the usual
$\delta\Phi_\alpha = e^a\partial_a\Phi_\alpha$ for any translation,
where $e^a$ denotes the direction of the translation,
and $L$ does not depend on $x^a$ explicitly,
then $\delta L = \partial_a(e^aL)$, thus the choice $K^a=e^a L$ can be made.
The Noether current is then $J^a=\Tc^{ab}e_b$ for any fixed $e^a$, where
$\Tc^{ab}=\frac{\partial L}{\partial(\partial_a\Phi_\alpha)}\partial^b\Phi_\alpha - \eta^{ab}L$.
Since $J^a$ is conserved for any $e^a$, 
$\Tc^{ab}$ itself is conserved,
i.e.\ $\partial_a \Tc^{ab}=0$ for the solutions of the Euler--Lagrange equations.
$\Tc^{ab}$ is called the \emph{canonical energy-momentum tensor}.
If $L$ is a total divergence, then the corresponding canonical energy-momentum tensor is obviously strongly conserved.
The canonical energy-momentum tensor depends on the choice of the Lagrangian for a given theory;
if the Lagrangian is modified by adding a total divergence, then the canonical energy-momentum tensor
changes by a strongly conserved term.

Trivially conserved tensors are often added to $\Tc^{ab}$ to obtain energy-momentum tensors
with better properties than $\Tc^{ab}$ (see \cite{BGRS}, for example).
The canonical energy-momentum tensor
of the electromagnetic field, for instance, is not symmetric and gauge invariant,
but by adding a trivially conserved tensor it is possible to obtain the
usual symmetric and gauge invariant energy-momentum tensor $T_{\mathrm{em}}^{ab}$.
In Section \ref{sec.emtensor} we apply the same correction method to obtain $T_{\mathrm{lg}}^{ab}$.

In electrodynamics the correction term to the canonical energy-momentum tensor
also follows from Noether's theorem---it is associated
with certain field dependent gauge transformations.
As a consequence, $T_{\mathrm{em}}^{ab}$ can be obtained by Noether's theorem directly \cite{BR},
\cite{BGRS}-\cite{MonFlo};
$T_{\mathrm{em}}^{ab}e_b$ is associated with the variation $\delta A_a=F_{ba}e^b$,
which corresponds to a translation accompanied by a field dependent gauge transformation.
In Section \ref{sec.emtensor} it will be seen that a similar result holds in linearized gravity as well.

In the following we shall use the letters $c$ and $d$, instead of $a$ and $b$,
in linearized gravity for the indices of currents and energy-momentum tensors,
as this turns out to be more convenient.

We note that in this paper we concentrate mostly on conserved currents rather than charges
and we do not use the Hamiltonian formalism and the action integral,
therefore we do not discuss boundary conditions or precisely specified fall off rates at infinity.

\subsection{Noether currents associated with gauge transformations}
\label{sec.gsnc}

{\bf Linearized gravity}\\
The variation of $L_1$, $L_2$ and $L_3$ under the action of a gauge transformation (\ref{eq.lg.g}) is 
\bea
\label{eq.lg.dl1}
\delta L_1 &  =  & \partial_c (\partial_a \phi_b F^{abc}) \\
\label{eq.lg.dl2}
\delta L_2 & = & 2\partial_c (\partial_a \phi_b\partial^a h^{bc} + \partial_b\phi_a \partial^a h^{bc}
-\partial^c\phi_a \partial_b h^{ba} - \partial_a\phi^c \partial_b h^{ba}) \\
\label{eq.lg.dl3}
\delta L_3 & = & \partial_c(\partial_b\phi_a\tilde{\mathring{F}}^{cab}).
\eea
To obtain (\ref{eq.lg.dl3}) we used $L_3=\frac{1}{2}\tilde{\mathring{F}}^{abc}\mathring{F}_{abc}$
and (\ref{eq.circ3}).
From (\ref{eq.lg.dl1}), (\ref{eq.lg.dl2}) and (\ref{eq.lg.dl3})
it can be seen that 
the gauge transformations (\ref{eq.lg.g}) are symmetries of $L$ with
\bea
K_{\mathrm{g}}^c & = & -\, \partial_a \phi_b F^{bac}
+ 2\alpha_2 (\partial_a \phi_b\partial^a h^{bc} + \partial_b\phi_a \partial^a h^{bc}
-\partial^c\phi_a \partial_b h^{ba} - \partial_a\phi^c \partial_b h^{ba}) \nonumber \\
&& +\, \alpha_3 \tilde{\mathring{F}}^{cab}\partial_b\phi_a
\label{eq.kg}
\eea
as $K^c$ in the identity $\delta L = \partial_c K^c$.
$K_{\mathrm{g}}^c$ is a reasonable choice for $K^c$, as it is linear in $\partial_a\phi_b$ and in $\partial_c h_{ab}$
and does not depend on higher derivatives of $h_{ab}$ and $\phi_a$.

Using
$\frac{\partial L}{\partial (\partial_c h_{ab})}  =  \frac{1}{2}(F^{cab}+F^{cba}) 
+ \alpha_2 (\partial^{a} h^{bc} + \partial^{b} h^{ac}
-\eta^{ca}\partial_d h^{bd} - \eta^{cb}\partial_d h^{ad} )
+ \frac{1}{2}\alpha_3 (\tilde{\mathring{F}}^{cab} + \tilde{\mathring{F}}^{cba})$,
the Noether current associated with a gauge transformation (\ref{eq.lg.g}) is then 
\bea
\label{eq.lg.jg00}
J^c_{\mathrm{g}} & = &  
2F^{cab} \partial_a \phi_b + \alpha_3 \tilde{\mathring{F}}^{cab} \partial_a\phi_b\\
& = & 2( \partial_a [F^{cab} \phi_b] -  \partial_a F^{cab}\phi_b)
+ \alpha_3 \partial_a [\tilde{\mathring{F}}^{cab} \phi_b].
\label{eq.lg.jg0}
\eea
In (\ref{eq.lg.jg0}) the terms $\partial_a [F^{cab} \phi_b]$ and $\partial_a [\tilde{\mathring{F}}^{cab} \phi_b]$ 
are strongly conserved, $\partial_a F^{cab}$ is weakly zero,
thus $J^c_{\mathrm{g}}$ is a trivially conserved current, as it has to be.
$J^c_{\mathrm{g}}$ does not depend on higher than first derivatives of $h_{ab}$,
it depends on $h_{ab}$ only through $F_{abc}$, it does not depend on higher than first derivatives of $\phi_a$,
it is a bilinear local function of $h_{ab}$ and $\phi_a$,
and it is independent of $\alpha_2$. 
In Section \ref{sec.emtensor} it will be connected with certain possible trivially conserved contributions
to the energy-momentum tensor of the linearized gravitational field.
Considering the corresponding current in electrodynamics (see (\ref{eq.emg2j}) below), 
$F^{cab} \partial_a \phi_b$ can be said to be an electric type current,
whereas $\tilde{\mathring{F}}^{cab} \partial_a\phi_b$ a magnetic type current.

Currents associated with gauge transformations can also be obtained from the
superpotential given in \cite{AS} (see equation (4.48)) or in \cite{AD,BB,CoOlSe},
but they differ from $J^c_{\mathrm{g}}$ somewhat; in particular,
they also depend on second derivatives of $\phi_a$ and on $h_{ab}$ without derivatives.

It must be noted that there is some arbitrariness in the above choice of $K^c$ even if only terms that are bilinear
in $\partial_a\phi_b$ and $\partial_ch_{ab}$ are allowed,
since under the latter condition it is still possible to add $\tilde{\mathring{F}}^{cab} \partial_a\phi_b$
to $K^c$ with an arbitrary coefficient. Such a modification of $K^c$ has the effect that the coefficient of
$\tilde{\mathring{F}}^{cab} \partial_a\phi_b$ changes in $J^c_{\mathrm{g}}$.

$J^c_{\mathrm{g}}$ can obviously be generalized to the case when matter is also present by
replacing $\partial_a F^{cab}$, which is weakly zero only if $\mc{T}^{ab}=0$,
with $\partial_a F^{cab}  - \mc{T}^{cb}$, which is weakly zero also in the presence of matter:
\bea
\label{eq.lg.jg00T}
J^c_{\mathrm{g,\mc{T}}} & = &  
2F^{cab} \partial_a \phi_b + 2\mc{T}^{cb}\phi_b  + \alpha_3 \tilde{\mathring{F}}^{cab} \partial_a\phi_b\\
& = & 2\bigl( \partial_a [F^{cab} \phi_b] -  ( \partial_a F^{cab} - \mc{T}^{cb}) \phi_b\bigr)
+ \alpha_3 \partial_a [\tilde{\mathring{F}}^{cab} \phi_b].
\label{eq.lg.jg0T}
\eea

One can say that a parameter $\phi_a$ of a gauge transformation is Killing vector-like if $\delta h_{ab}=\partial_a\phi_b+\partial_b\phi_a=0$.
The Killing vector-like $\phi_a$ parameters evidently coincide with the proper Killing vector fields of the Minkowski spacetime.
The two main subtypes of these Killing vector fields are the spacetime translation generators,
which are constant vector fields, and the Lorentz transformation generators,
which have the form $\phi_a = \omega_{ab}x^b$, where $\omega_{ab}$ is an antisymmetric tensor.
In the first case $J^c_{\mathrm{g,\mc{T}}} = 2\mc{T}^{cb}\phi_b$, in the second case
$J^c_{\mathrm{g,\mc{T}}} = F^{abc}\omega_{ab} + 2\mc{T}^{cb}\phi_b - \alpha_3 \tilde{\mathring{F}}^{cab}\omega_{ab}$.
The first two terms in the latter current are conserved separately as well.

The result $J^c_{\mathrm{g,\mc{T}}} = 2\mc{T}^{cb}\phi_b$ in the first case
shows that (i) $\mc{T}^{ab}$ can be regarded as a conserved tensor field
corresponding to particular gauge transformations, namely to the spacetime translation-like gauge transformations
(characterized by constant $\phi_a$),
and (ii) the linearized gravitational field does not carry any charge related to these gauge transformations.
$\mc{T}^{ab}$ and the spacetime translation-like gauge transformations are analogous to the
electric current and to the global gauge transformations
in electrodynamics---see the remark below (\ref{eq.emg2j}).\\

%
% ------------------------------------------------------------------------- 
%

\noi
{\bf Electrodynamics}\\
The analogous currents in electrodynamics in the absence of electric current are 
\beq
\label{eq.emg2}
J_{\mathrm{g,em}}^a = -F^{ab}\partial_b\phi  + \alpha_3 \tilde{F}^{ab}\partial_b\phi 
= -\partial_b [F^{ab}\phi ] + \partial_b F^{ab} \phi
+ \alpha_3\partial_b [\tilde{F}^{ab}\phi],
\eeq
where $\phi$ is an arbitrary scalar function specifying a gauge transformation, under which
the variation of $A_a$ is $\delta A_a = \partial_a \phi$.
$\partial_b [F^{ab}\phi]$ and $\partial_b [\tilde{F}^{ab}\phi]$ are strongly conserved, 
$\partial_b F^{ab}\phi$ is weakly zero, thus $J_{\mathrm{g,em}}^a$ is also a trivially conserved current.
The generalization of $J_{\mathrm{g,em}}^a$ to the case when the electric current is not zero,
obtained by replacing $\partial_b F^{ab}$ with $\partial_b F^{ab} + \mc{J}^a$, is
\beq
\label{eq.emg2j}
J_{\mathrm{g,\mc{J},em}}^a = -F^{ab}\partial_b\phi + \mc{J}^a\phi + \alpha_3 \tilde{F}^{ab}\partial_b\phi 
= -\partial_b [F^{ab}\phi ] + (\partial_b F^{ab} + \mc{J}^a)\phi
+ \alpha_3\partial_b [\tilde{F}^{ab}\phi].
\eeq
It is worth noting that if $\phi=1$, then $J_{\mathrm{g,\mc{J},em}}^a = \mc{J}^a$ and $\delta A_a = 0$.

(\ref{eq.emg2}) can be derived in the following way: 
The most general Lagrangian density function that is Lorentz covariant,  
differs from the standard Lagrangian $-\frac{1}{4}F_{ab}F^{ab}$ only by a total divergence and 
is a homogeneous quadratic expression in $\partial_a A_b$ with constant coefficients is 
\beq
\label{eq.eml}
L_{\mathrm{em}}=L_{1,\mathrm{em}} + \alpha_2 L_{2,\mathrm{em}} + \alpha_3 L_{3,\mathrm{em}},
\eeq
where $\alpha_2$ and $\alpha_3$ are arbitrary constants and
\beq
\label{eq.eml2}
L_{1,\mathrm{em}}=-\frac{1}{4}F_{ab}F^{ab},\qquad
L_{2,\mathrm{em}} = \partial_a A_b \partial^b A^a - \partial_a A^a \partial_b A^b , \qquad
L_{3,\mathrm{em}}=\frac{1}{4}F_{ab}\tilde{F}^{ab}.
\eeq
$L_{2,\mathrm{em}}$ and $L_{3,\mathrm{em}}$ are total divergences:
\beq
\label{eq.eml3}
L_{2,\mathrm{em}}= \partial_a (A^b \partial_b A^a - A^a \partial_b A^b),\qquad
L_{3,\mathrm{em}} = \frac{1}{2}\partial_a (\epsilon^{abcd} A_b \partial_c A_d).
\eeq
Gauge transformations are symmetries of $L_{\mathrm{em}}$:
\beq
\label{eq.emg1}
\delta L_{\mathrm{em}} = 2\alpha_2 \partial_a (\partial_b \phi \partial^b A^a
-\partial^a\phi\partial_b A^b)\qquad (\delta L_{1,\mathrm{em}} = \delta L_{3,\mathrm{em}} = 0).
\eeq
(\ref{eq.emg1}) implies that one can choose $K^a$ to be
$K_{\mathrm{g,em}}^a = 2\alpha_2 (\partial_b \phi \partial^b A^a
-\partial^a\phi\partial_b A^b)$.
The Noether currents associated with gauge transformations are found then
to be the currents given by (\ref{eq.emg2}).
These currents are well known;
see \cite{AS, Seraj}, for example, for recent articles discussing them.
$F^{ab}\partial_b\phi$ and $\tilde{F}^{ab}\partial_b\phi$
are known as electric and magnetic type currents, respectively.

\section{Energy-momentum tensor of the linearized gravitational field}
\label{sec.emtensor}

The main aim in this section is to find a gravitational counterpart of the standard energy-momentum tensor
of the electromagnetic field.
In addition, the $4$-parameter family of energy-momentum tensors found in \cite{BS} without gauge fixing
is written in a more compact form,
its interpretation from the perspective of symmetries is elaborated on,
it is extended into a $6$-parameter family,
and the extension of these energy-momentum tensors to the case when matter is present is discussed.
It is also shown that the energy-momentum tensors
change under gauge transformations only by a trivially conserved tensor field. 
Throughout the present and subsequent sections,
whenever the value of $\mc{T}^{ab}$ has relevance, $\mc{T}^{ab}=0$ will be assumed,
if not indicated otherwise.

\subsection{Gravitational analogue of the energy-momentum tensor of the electromagnetic field}
\label{sec.tst}

The canonical energy-momentum tensor (see the first part of Section \ref{sec.gauge}) following from $L_1$ is
\beq
\label{eq.Tc}
\Tc^{cd} = \frac{\partial L_1}{\partial (\partial_c h_{ab})}\partial^d h_{ab} - \eta^{cd}L_1 
= F^{cab}\partial^d h_{ab}
- \frac{1}{2} \eta^{cd} F^{eab}\mathring{F}_{eab}.
\eeq
$\Tc^{cd}$ is not scalar gauge invariant and hence cannot be expressed in terms of $F_{abc}$.
Nevertheless, it is possible to obtain a scalar gauge invariant energy-momentum tensor
by subtracting the trivially conserved tensor $F^{cab}\partial_ah^d{}_b$ from $\Tc^{cd}$,
in analogy with electrodynamics, where the standard gauge invariant energy-momentum tensor\\
$T_{\mathrm{em}}^{cd} = - F^{ca}F^d{}_a + \frac{1}{4}\eta^{cd}F_{ab}F^{ab}$
can be obtained by a similar correction of the canonical energy-momentum tensor. 
(In electrodynamics the correction term reads $F^{ca}\partial_aA^d$.
The fact that $F^{cab}\partial_ah^d{}_b$ is trivially conserved is shown by the identity
$F^{cab}\partial_ah_b{}^d = \partial_a (F^{cab}h_b{}^d) - \partial_a F^{cab}h_b{}^d$,
where the first term on the right hand side is obviously strongly conserved
and the second term is weakly zero in the absence of matter.)
The corrected energy-momentum tensor obtained in this way is
\beq
\label{eq.lg.Tst}
T_{\mathrm{lg}}^{cd} = 2\left(F^{cab}\mathring{F}^d{}_{ab}
- \frac{1}{4} \eta^{cd} F^{eab}\mathring{F}_{eab}\right),
\eeq
which is obviously scalar gauge invariant.
It is also traceless and its form shows good analogy with the form of the standard energy-momentum tensor
of the electromagnetic field,
especially if the first order form (\ref{eq.lg.foe}) of the gravitational field equations is considered. 
On the other hand, it is generally not symmetric. 
In Section \ref{sec.harmonic} it will be seen, nevertheless,
that $T_{\mathrm{lg}}^{cd}$ becomes symmetric and
satisfies the dominant energy condition in suitable gauges,
including the transverse traceless gauge.

As was mentioned in Section \ref{sec.gauge}, $T_{\mathrm{em}}^{cd}$ can also be obtained by Noether's theorem directly,
i.e.\ not only by adding a trivially conserved term to the canonical energy-momentum tensor.
A similar result, presented in Section \ref{sec.4p}, holds for $T_{\mathrm{lg}}^{cd}$ as well.

The divergence of the energy-momentum tensor in the presence of matter has an important role in
\cite{BHL1,BS}, therefore we write down expressions for $\partial_c T_{\mathrm{lg}}^{cd}$
and $\partial_c \Tc^{cd}$:
\beq
\label{eq.lg.dT}
\partial_c T_{\mathrm{lg}}^{cd} = - 2 \mc{T}^{ab}\mathring{F}^d{}_{ab},\qquad
\partial_c \Tc^{cd} = - \mc{T}^{ab}\partial^d h_{ab}.
\eeq
$\partial_c \Tc^{cd}$ has the form preferred in \cite{BHL1} (see equation (15)), up to a normalization factor $1/2$.
These equations can be regarded as local balance equations quantifying the
nonconservation of the energy and momentum of the linearized gravitational field due to the presence of matter.
The corresponding equations in electrodynamics are
\beq
\label{eq.em.ll1}
\partial_c T_{\mathrm{em}}^{cd} = -\mc{J}^aF^d{}_a,\qquad
\partial_c \Tc_{\mathrm{em}}^{cd}=-\mc{J}^a \partial^d A_a.
\eeq
The equations in (\ref{eq.lg.dT}) do not require $T_{\mathrm{lg}}^{cd}$ and $\Tc^{cd}$ to be gauge invariant,
because $\mc{T}^{ab}$ is multiplied by gauge dependent factors on the right hand sides.

Both $\partial_c T_{\mathrm{lg}}^{cd}$ and $\partial_c \Tc^{cd}$ have the form required in \cite{BS},
therefore both $T_{\mathrm{lg}}^{cd}$ and $\Tc^{cd}$ are necessarily included in the $4$-parameter family of
energy-momentum tensors found in \cite{BS} (see equation (23) of \cite{BS}).
In this sense $T_{\mathrm{lg}}^{cd}$ is not entirely new,
nevertheless it is not singled out in \cite{BS}. 
The methods used in \cite{BS} to obtain the energy-momentum tensors are also different from the approach
taken here.

\subsection{The $\mathbf{4}$-parameter family of energy-momentum tensors}
\label{sec.4p}

As explained below, the $4$-parameter family of energy-momentum tensors found in \cite{BS} can be written in the form
\beq
T^{cd}= \lambda_1 T_{\mathrm{lg}}^{cd}
+ \lambda_2 \partial_a U^{cad}_2
+ \lambda_3 F^{cab}\partial_a h_b{}^d +  \lambda_4 F^{cad}\partial_a h , 
\label{eq.lg.em}
\eeq
where $\lambda_1,\dots,\lambda_4$ are arbitrary constant coefficients and
\beq
\label{eq.lg.u2}
U_2^{cad} = [h^{bc}\partial^d h_b{}^a -  h^{ba}\partial^d h_b{}^c]
+ [\partial^b h^{ec} h_{be}\eta^{ad} - \partial^b h^{ea} h_{be}\eta^{cd} ]
+ [\partial_b h^{be} h^a{}_e \eta^{cd} - \partial_b h^{be} h^c{}_e \eta^{ad} ].
\eeq
$U_2^{cad}$ is antisymmetric in the first two indices, therefore $\partial_a U^{cad}_2$ is strongly conserved.
The terms multiplied by $\lambda_3$ and $\lambda_4$ are also trivially conserved in the absence of matter;
for $F^{cab}\partial_ah_b{}^d$ this has been shown above, and for $F^{cad}\partial_a h$ it can be seen from the identity 
$F^{cad}\partial_a h = \partial_a(F^{cad}h) - \partial_a F^{cad} h$,
where on the right hand side the first term is strongly conserved and the second term is weakly zero.
The tensors (\ref{eq.lg.em}) thus belong to a single equivalence class
(according to the definition of equivalence in Section \ref{sec.gauge}) for any fixed value of $\lambda_1$,
and give the same total energy and momentum (up to the normalization factor $\lambda_1$)
in the absence of matter
if $h_{ab}$ and $\partial_c h_{ab}$ fall off sufficiently rapidly at infinity.
Since the canonical energy-momentum tensor (\ref{eq.Tc})
is contained by the family (\ref{eq.lg.em}),
it also follows that the entire family (\ref{eq.lg.em}) can be said to be associated with spacetime translations
and its members can indeed be regarded as energy-momentum tensors.

The above relation of the family (\ref{eq.lg.em}) to spacetime translations was established also in \cite{BS}:
in the absence of matter
the tensors given by equation (23) of \cite{BS} were shown
to constitute a single equivalence class (up to normalization) that
contains the canonical energy-momentum tensor following from the Fierz--Pauli Lagrangian.
Below we supplement this with a more detailed interpretation of (\ref{eq.lg.em}) from the perspective of symmetries.

First, we note that it is possible to attach specific meaning
to all four terms on the right hand side of (\ref{eq.lg.em}) from the perspective of symmetries.
Regarding $\partial_a U^{cad}_2$, 
it is not difficult to verify that
it is the canonical energy-momentum tensor following from $L_2$.
The terms $F^{cab}\partial_a h_b{}^d$ and $F^{cad}\partial_a h$, on the other hand, can be related to gauge symmetry:
by substituting $\phi_b$ with $h_b{}^de_d$, where $e_d$ is an arbitrary constant vector,
in the current $F^{cab}\partial_a\phi_b$ (see (\ref{eq.lg.jg00})), we get $F^{cab}\partial_a h_b{}^de_d$,
whereas by substituting $\phi_b$ with $he_b$, we get $F^{cad}\partial_a h e_d$.
This also means that $F^{cab}\partial_a h_b{}^de_d$ and $F^{cad}\partial_a h e_d$ are Noether currents associated with
the variations
$\delta h_{ab}=\frac{1}{2}(\partial_a(h_b{}^de_d)+\partial_b(h_a{}^de_d))$
and
$\delta h_{ab}=\frac{1}{2}(\partial_a(he_b)+\partial_b(he_a))$, respectively.
These variations correspond to field dependent (generalized) gauge transformations.
For $F^{cab}\partial_a h_b{}^de_d$ and $F^{cad}\partial_a h e_d$,
the relevant $K^c$ appearing in the identity $\delta L_1=\partial_cK^c$
is $\frac{1}{2}F^{bac}\mathring{F}_{bad}e^d$ and $\frac{1}{2}F^{abc}\partial_ah e_b$, respectively,
as obtained from $K_{\mathrm{g}}^c|_{\alpha_2=\alpha_3=0}$ (see (\ref{eq.kg})) by the replacements $\phi_a=\frac{1}{2}h_{ab}e^b$
and $\phi_a=\frac{1}{2}he_a$.
Remarkably, in electrodynamics there is only one similar current (for fixed $e_d$), namely $F^{ca}\partial_a A^de_d$,
which can be obtained from $F^{ca}\partial_a\phi$ (see (\ref{eq.emg2}))
by the replacement $\phi=A^de_d$.

From the considerations in the previous paragraphs and the additivity of Noether currents it follows that
$T^{cd}e_d$ (with $\lambda_1=1$) is a Noether current, associated with the symmetry of $L_1$ characterized by the variation
\bea
\delta h_{ab} & = &  e^c\partial_ch_{ab} + \frac{1}{2}(\lambda_3 - 1)\delta_{\mathrm{g}}h_{ab}|_{\phi_a=h_{ab}e^b}
+ \frac{1}{2}\lambda_4 \delta_{\mathrm{g}}h_{ab}|_{\phi_a=he_a}\\
& = & (\mathring{F}_{cab}+\mathring{F}_{cba})e^c
+ \frac{\lambda_3}{2}(e^c\partial_ah_{bc}+e^c\partial_bh_{ac})
+ \frac{\lambda_4}{2}(e_b\partial_a h + e_a\partial_bh),
\label{eq.lg.mdh}
\eea
where $\delta_{\mathrm{g}}h_{ab}=\partial_a\phi_b + \partial_b\phi_a$.
This $\delta h_{ab}$ corresponds to a translation accompanied by a gauge transformation,
although the gauge transformation is understood here in a generalized sense, as its parameter depends on $h_{ab}$.
For $\Tc^{cd}e_d$, $K^c$ is $e^cL_1$,
thus for $T^{cd}e_d$, $K^c$ is 
\beq
K^c=e^cL_1  
-\lambda_2\partial_a U_2^{cad}e_d
+\frac{1}{2}(\lambda_3-1)F^{bac}\mathring{F}_{bad}e^d
+\frac{1}{2}\lambda_4F^{abc}\partial_ah e_b.
\eeq

For $T_{\mathrm{lg}}^{cd}e_d$, in particular,
$\delta h_{ab}= (\mathring{F}_{cab}+\mathring{F}_{cba})e^c$
and $K^c=e^cL_1 - \frac{1}{2}F^{bac}\mathring{F}_{bad}e^d$.
This case can be compared to the case of the standard energy-momentum tensor of the electromagnetic field:
$T_{\mathrm{em}}^{cd}e_d$ is a Noether current associated with the variation $\delta A_a=F_{da}e^d$
with $K^c = e^cL_{1,\mathrm{em}}$. In contrast with linearized gravity,
the contribution to the latter $K^c$ from the generalized gauge transformation included in $\delta A_a=F_{da}e^d$ is zero,
because $L_{1,\mathrm{em}}$ is strictly gauge invariant.\\

Turning to the comparison of (\ref{eq.lg.em}) with the expression for this $4$-parameter family of energy-momentum tensors presented in \cite{BS},
we first note that $U_2^{cad}$ is, up to a normalization factor, the superpotential given in equation (22) of \cite{BS},
and $\partial_a U^{cad}_2$ is the term multiplied by $\alpha_2$ in equation (23) of \cite{BS}.
Similarly, the terms $F^{cad}\partial_a h$ and $F^{cab}\partial_a h_b{}^d$
agree with the terms multiplied by $\alpha_3$ and $\alpha_4$ in equation (23) of \cite{BS}.
$T_{\mathrm{lg}}^{cd}$ differs from the tensor multiplied by $\alpha_1$ in equation (23) of \cite{BS},
but we have seen that it is a member of the $4$-parameter family given by that equation.
Furthermore, it is not a linear combination of $\partial_a U^{cad}_2$, $F^{cad}\partial_a h$ and $F^{cab}\partial_a h_b{}^d$,
because its trace is zero, whereas
the linear combinations of $\partial_a U^{cad}_2$, $F^{cad}\partial_a h$ and $F^{cab}\partial_a h_b{}^d$
are not traceless, as can be verified by direct calculation.
This means that $T_{\mathrm{lg}}^{cd}$ indeed spans a $4$-dimensional family of tensors together with
$\partial_a U^{cad}_2$, $F^{cad}\partial_a h$ and $F^{cab}\partial_a h_b{}^d$,
and this $4$-parameter family is the same as the one given by equation (23) of \cite{BS}.\\

\noi
{\bf Symmetric energy-momentum tensor}\\
According to the results of \cite{BS},
the family (\ref{eq.lg.em}) contains a single symmetric tensor (assuming $\lambda_1=1$), 
which is the linearized Landau--Lifshitz pseudotensor $T_{\mathrm{LL}}^{cd}$
multiplied by $2$.
It corresponds to $\lambda_1=1$, $\lambda_2=-\frac{1}{4}$, $\lambda_3=-1$, $\lambda_4=1$
in terms of our parameters.
By direct calculation it can be seen that $T_{\mathrm{LL}}^{cd}$ is not
scalar gauge invariant and its trace is not zero.
$T_{\mathrm{lg}}^{cd}$ is the only traceless member of the family (\ref{eq.lg.em}),
and it is also the only member that is scalar gauge invariant. \\

\noi
{\bf $\mathbf{6}$-parameter extension}\\
$\tilde{\mathring{F}}^{cab}\partial_ah_b{}^d$ and $\tilde{\mathring{F}}^{cad}\partial_ah$
are further possible correction terms that can
be included in the general formula for $T^{cd}$.
These terms contain $\epsilon^{abcd}$ and were apparently not considered in \cite{BS}.
They can be obtained by substituting $\phi_b$ with $h_b{}^de_d$ or $he_b$ in the strongly conserved current
$\tilde{\mathring{F}}^{cab}\partial_a\phi_b$ (see (\ref{eq.lg.jg00})).
A third term containing $\epsilon^{abcd}$ would be the canonical energy-momentum tensor corresponding to $L_3$,
but this turns out to coincide with $\tilde{\mathring{F}}^{cab}\partial_ah_b{}^d$.
$\tilde{\mathring{F}}^{cab}\partial_ah_b{}^d$ and $\tilde{\mathring{F}}^{cad}\partial_ah$ are strongly conserved tensor fields,
as the identities $\tilde{\mathring{F}}^{cab}\partial_ah_b{}^d = \partial_a(\tilde{\mathring{F}}^{cab}h_b{}^d)$,
$\tilde{\mathring{F}}^{cad}\partial_ah = \partial_a(\tilde{\mathring{F}}^{cad}h)$ show.
$\tilde{\mathring{F}}^{cad}\partial_ah$ is traceless (see (\ref{eq.lg.x2b})), but not scalar gauge invariant.\\

\noi
{\bf Extension in the presence of matter}\\
In \cite{BHL1}, the balance equation describing the nonconservation of the energy and momentum
of the linearized gravitational field in the presence of matter is required to take the form (see equation (15))
\beq
\label{eq.Tbal}
\partial_c T^{cd} = - \frac{1}{2}\mc{T}^{ab}\partial^d h_{ab}.
\eeq
As was mentioned above, the canonical energy-momentum tensor $\Tc^{cd}$ satisfies this equation
up to a normalization factor, but $T_{\mathrm{lg}}^{cd}$ and generally (\ref{eq.lg.em}) do not.
(\ref{eq.Tbal}) is not required from energy-momentum tensors in the approach taken in this paper,
nevertheless it is worth mentioning that
it is possible to extend $T_{\mathrm{lg}}^{cd}$ and (\ref{eq.lg.em}) so that they satisfy (\ref{eq.Tbal}) up to normalization.
For $T_{\mathrm{lg}}^{cd}$, a possible extended form is
$T_{\mathrm{lg}}^{cd} - \mc{T}^{cb}h_b{}^d$, which is obtained by subtracting
$F^{cab}\partial_ah^d{}_b + \mc{T}^{cb}h_b{}^d$, instead of $F^{cab}\partial_ah^d{}_b$,
from $\Tc^{cd}$. The reason for this modification of the correction term $F^{cab}\partial_ah^d{}_b$
is that in the presence of matter $F^{cab}\partial_ah^d{}_b$ is not trivially conserved,
but $F^{cab}\partial_ah^d{}_b + \mc{T}^{cb}h_b{}^d$ is, as can be seen from the identity
$F^{cab}\partial_ah^d{}_b = \partial_a (F^{cab}h^d{}_b) - \partial_a F^{cab}h^d{}_b$.
To obtain the similar extended form of (\ref{eq.lg.em}), one also replaces $F^{cad}\partial_a h$
with $F^{cad}\partial_a h + \mc{T}^{cd}h$, which is trivially conserved in the presence of matter.
Strongly conserved tensors do not need modification.
The extended form of (\ref{eq.lg.em}) is then
\beq
T_{\mc{T}}^{cd} = \lambda_1 (T_{\mathrm{lg}}^{cd} - \mc{T}^{cb}h_b{}^d)
+ \lambda_2 \partial_a U^{cad}_2
+ \lambda_3 (F^{cab}\partial_a h_b{}^d + \mc{T}^{cb}h_b{}^d)
+ \lambda_4 (F^{cad}\partial_a h + \mc{T}^{cd}h).
\label{eq.lg.emmat}
\eeq
The balance equation takes the form $\partial_c T_{\mc{T}}^{cd} = - \lambda_1\mc{T}^{ab}\partial^d h_{ab}$
for $T_{\mc{T}}^{cd}$.
The terms in (\ref{eq.lg.emmat}) containing $\mc{T}^{ab}$ explicitly may be interpreted as interaction terms.

In contrast with (\ref{eq.lg.em}), the tensors (\ref{eq.lg.emmat})
constitute a single equivalence class for any fixed value of $\lambda_1$ also in the presence of matter,
thus the total energy and momentum yielded by $T_{\mc{T}}^{cd}$ for the linearized gravitational field
at any time is independent of $\lambda_2$, $\lambda_3$, $\lambda_4$ if $h_{ab}$ and $\partial_c h_{ab}$
fall off sufficiently rapidly at infinity.

\subsection{Gauge invariance of the energy-momentum tensors up to a trivially conserved tensor}
\label{sec.ginvemt}

In \cite{Wald1990} it is shown that it is a general feature of conserved currents in gauge theories
that they change by a trivially conserved current under gauge transformations.
Nevertheless, we present here some details of the derivation of this remarkable invariance property
for $T_{\mathrm{lg}}^{cd}$
and for the other energy-momentum tensors differing from $T_{\mathrm{lg}}^{cd}$
by a trivially conserved term.
We assume $\mc{T}^{cd}=0$, since $T_{\mathrm{lg}}^{cd}$ and the other energy-momentum tensors are not conserved otherwise.

By taking the first order variation of the identity
$\partial_c T_{\mathrm{lg}}^{cd} = -2G^{ab}\mathring{F}^d{}_{ab}$ with respect to a gauge transformation
and using
$\delta G^{ab}=0$ and $\delta\mathring{F}^d{}_{ab} = \frac{1}{2}(\partial^d{}_b\phi_a - \partial_{ab}\phi^d)$,
we get the identity
$\partial_c \delta T_{\mathrm{lg}}^{cd} = -2G^{ab}\delta\mathring{F}^d{}_{ab}
= -G^{ab}(\partial^d{}_b\phi_a - \partial_{ab}\phi^d)$
(as before, $\delta$ indicates first order variation).
Taking into account $\partial_bG^{ab}=0$, this identity can be written as
\beq
\label{eq.gsym}
\partial_c ( \delta T_{\mathrm{lg}}^{cd} + G^{ac}\partial^d\phi_a - G^{ac}\partial_a\phi^d ) = 0.
\eeq
$G^{ac}\partial^d\phi_a - G^{ac}\partial_a\phi^d$ is weakly zero in the absence of matter,
thus (\ref{eq.gsym}) means that $\delta T_{\mathrm{lg}}^{cd}$ is a trivially conserved tensor.
A superpotential $\Sigma^{ced}$ for $\delta T_{\mathrm{lg}}^{cd} + G^{ac}\partial^d\phi_a - G^{ac}\partial_a\phi^d$
can be computed by applying a general theorem that can be found in \cite{Fletcher} (Section VI) or in
somewhat different form in \cite{Wald1990} 
(it is quoted also in \cite{TG2}; see equations (2.33)-(2.36) in Section II.D).
$\Sigma^{ced}$ is a homogeneous bilinear local function of $h_{ab}$ and $\phi_a$.
The full (i.e., not only first order) variation of $T_{\mathrm{lg}}^{cd}$ under a gauge transformation is
$\delta^{(2)} T_{\mathrm{lg}}^{cd} + \delta T_{\mathrm{lg}}^{cd}$,
where $\delta^{(2)} T_{\mathrm{lg}}^{cd} = 
\frac{1}{2}(\delta T_{\mathrm{lg}}^{cd})|_{h_{ab}=\partial_a\phi_b+\partial_b\phi_a}$.
$( \delta T_{\mathrm{lg}}^{cd} + G^{ac}\partial^d\phi_a - G^{ac}\partial_a\phi^d )|_{h_{ab}=\partial_a\phi_b+\partial_b\phi_a}
=(\delta T_{\mathrm{lg}}^{cd})|_{h_{ab}=\partial_a\phi_b+\partial_b\phi_a}$,
since $G^{cd}|_{h_{ab}=\partial_a\phi_b+\partial_b\phi_a}=0$,
therefore 
$(\delta T_{\mathrm{lg}}^{cd})|_{h_{ab}=\partial_a\phi_b+\partial_b\phi_a}
= \partial_e\Sigma^{ced}|_{h_{ab}=\partial_a\phi_b+\partial_b\phi_a}$.
For the full variation of $T_{\mathrm{lg}}^{cd}$ we thus have the identity
\beq
\delta^{(2)} T_{\mathrm{lg}}^{cd} + \delta T_{\mathrm{lg}}^{cd} =
-( G^{ac}\partial^d\phi_a - G^{ac}\partial_a\phi^d )
+\partial_e (\Sigma^{ced} + \Sigma^{(2)ced}),
\label{eq.tfullv}
\eeq
where $\Sigma^{(2)ced} = \frac{1}{2}\Sigma^{ced}|_{h_{ab}=\partial_a\phi_b+\partial_b\phi_a}$.
$( G^{ac}\partial^d\phi_a - G^{ac}\partial_a\phi^d )$ is weakly zero and
$\partial_e (\Sigma^{ced} + \Sigma^{(2)ced})$
is strongly conserved (it has the superpotential $\Sigma^{ced} + \Sigma^{(2)ced}$),
thus $\delta^{(2)} T_{\mathrm{lg}}^{cd} + \delta T_{\mathrm{lg}}^{cd}$
is trivially conserved.

From (\ref{eq.tfullv}) it follows that
the total energy and momentum given by $T_{\mathrm{lg}}^{cd}$ is invariant
under any gauge transformation whose
parameter and its derivatives up to sufficiently high order
fall off sufficiently rapidly at spatial infinity.

It is easy to see that the variation of a trivially conserved current under a gauge transformation is also
a trivially conserved current.
This implies that not only $T_{\mathrm{lg}}^{cd}$, but all members of the $6$-parameter family of energy-momentum tensors
mentioned in Section \ref{sec.4p} are gauge invariant up to a trivially conserved tensor.
The total energy given by the members of this $6$-parameter family
also has the same gauge invariance property as the total energy given by $T_{\mathrm{lg}}^{cd}$.

\section{Gauge fixing}
\label{sec.harmonic}

In this section we continue the discussion of the energy-momentum tensor of the linearized gravitational field
and the form of the field equations for $F_{abc}$ in particular gauges.
We show that the field equations and $T_{\mathrm{lg}}^{cd}$ acquire additional favourable properties in suitable gauges.
We discuss the $3$-parameter family of symmetric energy-momentum tensors found in \cite{BS} and
make a few observations about $\tau^{cd}$.

We consider the generalized harmonic gauges,
the traceless harmonic gauge and the transverse traceless gauge,
which are well-known gauges for linearized gravity.
We also introduce some gauges defined by conditions that can be expressed in terms of the Fierz tensor. 
The latter gauges, which are well suited to the Fierz formalism, are
(a) $F_a=0$,
(b) $F_a=0$, $F_{ab0}=0$,
(c) $\partial_aF_b-\partial_bF_a=0$.
These gauges have close relations to the previously mentioned ones.
The $\partial_aF_b-\partial_bF_a=0$ gauge contains all generalized harmonic gauges,
the gauge $F_a=0$ is a special generalized harmonic gauge,
and the gauge $F_a=0$, $F_{ab0}=0$ is, in some cases, a natural alternative to the TT gauge,
which is contained by it.\\

\noi
{\bf The generalized harmonic gauges}

\noi
The generalized harmonic gauge condition for linearized gravity is
\beq
X^b(\chi) = \partial_a (h^{ab}-\chi \eta^{ab}h) = 0,
\eeq
where $\chi$ is a constant parameter of the gauge and can take arbitrary value.
The usual harmonic gauge corresponds to $\chi=\frac{1}{2}$. 
Another special case is $\chi=1$, when the gauge fixing condition becomes $F_a=0$.
Generally
\beq
\label{eq.ghgF}
F_a=(\chi-1)\partial_a h
\eeq
in generalized harmonic gauge.

It should be noted that the $F_a=0$ gauge is less general than the other generalized harmonic gauges,
since it requires $\mc{T}=0$, as can be seen by taking the trace of
the linearized Einstein equations $\partial_c F^{cab}= - \mc{T}^{ab}$.
The $F_a=0$ gauge is also the only generalized harmonic gauge that is preserved by scalar gauge transformations,
and can be regarded as the generalized harmonic gauge distinguished by its compatibility with the Fierz formalism.

$F_{abc}$ can be written as
\beq
\label{eq.lg.fstrength}
F_{abc}=\frac{1}{2}(\partial_a\bar{h}_{bc}-\partial_b\bar{h}_{ac})
\eeq
in generalized harmonic gauges, where
\beq
\bar{h}_{ab}=h_{ab}+(\chi-1)h\eta_{ab}.
\eeq

(\ref{eq.ghgF}) implies that
\beq
\label{eq.curlF}
\partial_aF_b-\partial_bF_a=0
\eeq
holds in any generalized harmonic gauge, i.e.\ the generalized harmonic gauges
are subgauges of the gauge determined by (\ref{eq.curlF}). It is also worth noting that
\beq
\label{eq.fx}
\partial_aF_b-\partial_bF_a=\partial_aX_b-\partial_bX_a.
\eeq

\noi
{\bf The transverse traceless gauge}

\noi
In the absence of matter it is possible to achieve both $\partial_a h^{ab}=0$ and $h=0$ by gauge transformation;
we call the gauge characterized by these conditions traceless harmonic gauge.
The traceless harmonic gauge is clearly a subgauge of all generalized harmonic gauges.

The traceless harmonic gauge conditions still leave some gauge freedom, which can be used to achieve
\beq
\label{eq.lg.tt1}
h_{i0} = 0,\quad (i=1,2,3),
\eeq
in addition to $\partial_a h^{ab}=0$ and $h=0$.
The gauge obtained in this way is the transverse traceless (TT) gauge.
If the behaviour of $h_{ab}$ is good at infinity,
then it is also possible to achieve
\beq
\label{eq.lg.tt2}
h_{a0}=0,
\eeq
instead of $h_{i0} = 0$ \cite{WaldGR}. (\ref{eq.lg.tt2}) can always be achieved in local vacuum regions as well \cite{FH}.
In many articles, including \cite{BHL1} and \cite{Barnett}, the TT gauge conditions are understood
to include the stronger condition (\ref{eq.lg.tt2}), therefore we shall do the same.\\

\noi
{\bf The $F_a=0$, $F_{ab0}=0$ gauge}

\noi
It is clear that in TT gauge $F_a=0$ and $F_{ab0}=0$. In the subsequent sections it will be seen that
in some contexts it is sufficient and natural to impose
only the latter conditions instead of the TT gauge conditions, which are more restrictive.

The conditions $F_a=0$ and $F_{ab0}=0$ fix the gauge completely up to scalar gauge transformations in the following sense:
\begin{lemma}
If $h_{ab}$ is defined on the entire Minkowski spacetime,
$\phi_a$ is the parameter of a gauge transformation that preserves the $F_a=0$, $F_{ab0}=0$ gauge,
and $\partial_a\phi_b-\partial_b\phi_a$ converges to zero at spatial infinity,
then $\phi_a=\partial_a\phi$ with some function $\phi$.
\end{lemma}
Proof. From the condition that $F_a=0$ is preserved, it follows that $\partial_a(\partial^a\phi^b-\partial^b\phi^a)=0$,
i.e.\ $\phi_a$ satisfies Maxwell's equations without source. If $F_{ab0}=0$ is also preserved,
then also $\partial_0(\partial_a\phi_b-\partial_b\phi_a)=0$.
Maxwell's equations imply that $\Box(\partial_a\phi_b-\partial_b\phi_a)=0$,
therefore from Maxwell's equations and $\partial_0(\partial_a\phi_b-\partial_b\phi_a)=0$ it follows
that $\partial_a\phi_b-\partial_b\phi_a$ satisfies the Laplace equation.
Together with the condition on the behaviour of $\partial_a\phi_b-\partial_b\phi_a$ at spatial infinity,
this implies that $\partial_a\phi_b-\partial_b\phi_a=0$.
Finally, from the last equation it follows that $\phi_a$ takes the form $\partial_a\phi$.
\hfill$\blacksquare$\\

Since $F_{abc}$ is scalar gauge invariant, any scalar gauge transformation preserves the conditions $F_a=0$, $F_{ab0}=0$.

\subsection{The field equations for $F_{abc}$}
\label{sec.ghgfieldeq}

It is not difficult to see that from (\ref{eq.curlF}) and from the definition (\ref{eq.trx1})
it follows that $\partial_c\tilde{\check{F}}^{cab}=0$,
therefore in the $\partial_aF_b-\partial_bF_a=0$ gauge
$\partial_c\tilde{F}^{cab}=0$ is equivalent with $\partial_c\tilde{\mathring{F}}^{cab}=0$.
This means that (\ref{eq.lg.foe}) is equivalent with the field equations  
\beq
\label{eq.lg.foeharm}
\partial_c F^{cab}= - \mc{T}^{ab},\qquad\quad \partial_c\tilde{F}^{cab}=0
\eeq
in the $\partial_aF_b-\partial_bF_a=0$ gauge.
These equations are clearly more symmetric than (\ref{eq.lg.foe}).
In generalized harmonic gauges $\partial_c\tilde{F}^{cab}=0$
can also be seen to follow directly from (\ref{eq.lg.fstrength}).

From (\ref{eq.lg.foeharm}) it follows that $F_{abc}$ satisfies the wave equation
\beq
\label{eq.lg.fwavharm}
\Box F_{abc} = - (\partial_a \mc{T}_{bc} - \partial_b\mc{T}_{ac})
\eeq
(the derivation is similar to the derivation of (\ref{eq.lg.fwav}) in Section \ref{sec.lg.wave}).
The form of the right hand side of (\ref{eq.lg.fwavharm}) is more desirable than that of (\ref{eq.lg.fwav}),
as $F_{abc}$ does not appear in it.

We also note that from (\ref{eq.eintr}) and (\ref{eq.ghgF}) it follows that
\beq
(\chi-1)\Box h = -\mc{T},
\eeq
i.e.\ $h$ satisfies the inhomogeneous wave equation in generalized harmonic gauge if $\chi\ne 1$.

\subsubsection{Gravitational analogues of the electric and magnetic fields}
\label{sec.eb}

In the $F_a=0$, $F_{ab0}=0$ gauge one can define 
the gravitational analogues $E_{ij}$, $B_{ij}$ of the electric and magnetic fields as
\beq
E_{ij} = -2F_{0ij},\qquad
B_{ij} = -2\tilde{F}_{0ij}.
\label{eq.eij}
\eeq
$F^{ijk}$ and $\tilde{F}^{ijk}$ can be expressed in terms of $B_{ij}$ and $E_{ij}$
as
\beq
F^{ijk}=\frac{1}{2}\epsilon^{ijl}B_l{}^k,\qquad \tilde{F}^{ijk}=-\frac{1}{2}\epsilon^{ijl}E_l{}^k.
\label{eq.eij2}
\eeq
(\ref{eq.eij}) and (\ref{eq.eij2}) show that $F_{abc}$ is completely determined by $E_{ij}$ and $B_{ij}$.

$E_{ij}$ is clearly traceless as a consequence of the gauge conditions.
From (\ref{eq.lg.x1}), (\ref{eq.lg.x2}) and $F_{ab0}=0$ it follows that $E_{ij}$ is symmetric.
It is also divergenceless, 
as a consequence of the identity $\partial_c F_{ab}{}^c=0$ and $F_{ab0}=0$.
$\tilde{F}_{abc}$ has the cyclic property as a consequence of $F_a=0$ (see (\ref{eq.lg.dualcycl})),
and $\tilde{F}_{ab0}=0$ follows from $F_{ab0}=0$. These properties of $\tilde{F}_{abc}$
and $\tilde{F}_{abc} = -\tilde{F}_{bac}$
imply that $B_{ij}$ is symmetric. From (\ref{eq.lg.x2b}) and $\tilde{F}_{ab0}=0$ it follows that
$B_{ij}$ is traceless. From the identity $\partial_c F_{ab}{}^c=0$ it follows that
$\partial_c \tilde{F}_{ab}{}^c=0$, and this implies, together with $\tilde{F}_{ab0}=0$,
that $B_{ij}$ is also divergenceless.

We now rewrite the field equations $\partial_c F^{cab} = 0$, $\partial_c \tilde{F}^{cab}=0$
in terms of $E_{ij}$ and $B_{ij}$.
Since $F^{ca0}=\tilde{F}^{ca0}=0$,
only the equations with $b=1,2,3$ are nontrivial. 
$\partial_c F^{cab} = 0$ can be written as
$\partial_0 F^{0ak}+\partial_i F^{iak}=0$, $k=1,2,3$. For $a=0$, this becomes
$\partial_i F^{i0k}=\frac{1}{2}\partial_i E^{ik}=0$, i.e.
\beq
\label{max1}
\partial_i E^{ik}=0.
\eeq
Similarly, $\partial_c\tilde{F}^{c0b}=0$ becomes
\beq
\label{max2}
\partial_i B^{ik}=0.
\eeq
As we have seen, (\ref{max1}) and (\ref{max2}) follow already from the gauge conditions.\\
For $a=1,2,3$, $\partial_c F^{cab} = 0$ becomes
\beq
\label{max4}
\partial_0 E^{ik} = -\epsilon^{ijl}\partial_j B_l{}^k.
\eeq
Finally, $\partial_c \tilde{F}^{cab} = 0$, $a=1,2,3$, becomes 
\beq
\label{max3}
\partial_0 B^{ik}=\epsilon^{ijl}\partial_j E_l{}^k.
\eeq
Equations (\ref{max1})-(\ref{max3}) are very similar to the vacuum Maxwell equations for the electric and magnetic fields.

We note that if $E_{ij}$ and $B_{ij}$ are symmetric and traceless fields, then $F_{abc}$, defined by
$F_{0ij}=-F_{i0j}=-\frac{1}{2}E_{ij}$, $F^{ijk}=-\frac{1}{2}\epsilon^{jil}B_l{}^k$,
$F_{ab0}=0$, $F_{00a}=0$, has the algebraic properties (\ref{eq.lg.x1}), (\ref{eq.lg.x2}), $F_a=0$,
and (\ref{eq.eij}) holds. If (\ref{max1})-(\ref{max3}) also hold,
then $\partial_c F^{cab} = 0$, $\partial_c \tilde{F}^{cab}=0$. 
This means that for symmetric and traceless fields $E_{ij}$ and $B_{ij}$,
equations (\ref{max1})-(\ref{max3}) are equivalent with
the equations
$\partial_c F^{cab} = 0$, $\partial_c \tilde{F}^{cab}=0$, $F_a=0$, $F_{ab0}=0$
for $F_{abc}$ (which is assumed to have the properties (\ref{eq.lg.x1}), (\ref{eq.lg.x2})).

In TT gauge 
\beq
E_{ij}=-\partial_0 h_{ij},\qquad
B_{ij}=-\epsilon_j{}^{lk} \partial_l h_{ki}.
\eeq
In \cite{Barnett}, where the TT gauge is used, $E_{ij}$ and $B_{ij}$ are defined by these equations.
The Maxwellian equations found in \cite{Barnett} (see equation (29)) for $E_{ij}$ and $B_{ij}$
agree with (\ref{max1})-(\ref{max3}).
Comparing with \cite{Barnett}, here we have found that these Maxwellian equations
remain valid in the gauge $F_a=0$, $F_{ab0}=0$ (which is an extension of the TT gauge),
and that they have a Lorentz covariant generalization, namely (\ref{eq.lg.foe}), which is valid without gauge fixing.

The value of $L_1$ for fields in the $F_a=0$, $F_{ab0}=0$ gauge is
$\frac{1}{2}(E_{ij}E^{ij} - B_{ij}B^{ij})$.
In TT gauge this coincides, up to normalization, with the expression given in equation (41) of \cite{Barnett}
for the Lagrangian density $\mc{L}'$ 
introduced in equation (40) of \cite{Barnett}.

Starting from the Fierz tensor, gravitoelectric and -magnetic vector fields
are introduced in \cite{ABN,UmGal1,UmGal2,UmGal3}; these fields are different from $E_{ij}$ and $B_{ij}$.
For an algebraic $(3+1)$-decomposition of $\breve{F}_{abc}$ without the condition $\breve{F}_{ab0}=0$
(which is also applicable to $F_{abc}$ if $F_a=0$),
see Appendix C of \cite{JWK}.

\subsection{Properties of $T_{\mathrm{lg}}^{cd}$}
\label{sec.tlgpos}

$T_{\mathrm{lg}}^{cd}$ is symmetric if $F_a=0$,
as $F_a=0$ implies $F_{abc}=\mathring{F}_{abc}$:
\beq
\label{eq.lg.Tst2}
T_{\mathrm{lg}}^{cd}|_{F_a=0} = 2\left(F^{cab}F^d{}_{ab}
- \frac{1}{4} \eta^{cd} F^{eab}F_{eab}\right).
\eeq
This expression resembles
the standard energy-momentum tensor of the electromagnetic field even more than (\ref{eq.lg.Tst}).
As $F_a=0$ in traceless harmonic gauge and in TT gauge, (\ref{eq.lg.Tst2}) is valid also in these gauges.

%%%

In TT gauge,
the energy density $T_{\mathrm{lg}}^{00}$, the momentum densities $T_{\mathrm{lg}}^{0k}$
and the stress tensor $T_{\mathrm{lg}}^{jk}$ can be written in terms of $E_{ij}$ and $B_{ij}$ as
\beq
\label{eq.ttt1}
T_{\mathrm{lg}}^{00}=\frac{1}{4} (E_{ij}E^{ij} + B_{ij}B^{ij}),\qquad 
T_{\mathrm{lg}}^{0k}=\frac{1}{2} \epsilon^{klj}E^i{}_l B_{ij},
\eeq
\beq
\label{eq.ttt2}
T_{\mathrm{lg}}^{jk}=\frac{1}{2}(E^{ji}E_i{}^k + B^{ji}B_i{}^k) - \frac{1}{4}\eta^{jk}(E_{lm}E^{lm}+B_{lm}B^{lm}),
\eeq
which are very similar to the corresponding expressions in electrodynamics. 
Up to normalization, the expressions in (\ref{eq.ttt1}) coincide with those found in \cite{Barnett} (see equations (43), (62)).
The energy density appears in \cite{AfSt} and as Hamiltonian density in \cite{DS2} as well.
The stress tensor was not considered in \cite{Barnett}.
(\ref{eq.ttt1}) and (\ref{eq.ttt2}) are already valid if $F_a=0$ and $F_{ab0}=0$, i.e.\
the TT gauge conditions are not necessary.

\subsubsection{Positivity properties of $T_{\mathrm{lg}}^{cd}$}

From (\ref{eq.ttt1}) it is clear that
the energy density $T_{\mathrm{lg}}^{00}$ is nonnegative if $F_a=0$, $F_{ab0}=0$. 
More generally, $T_{\mathrm{lg}}^{cd}$ has the following positivity properties:
\begin{prop}
\label{tstpos}
If $F_a=0$ and $F_{ab0}=0$, then 
$T_{\mathrm{lg}}^{cd}$ satisfies the dominant energy condition,
i.e.\ $T_{\mathrm{lg}}^{cd}v_c w_d\ge 0$ for any future-pointing time-like or null vectors $v^a$, $w^a$.
If $F_a=0$, $F_{ab0}=0$ and $T_{\mathrm{lg}}^{00}=0$
on a spacetime domain $\Omega$ with $H_{\mathrm{dR}}^1(\Omega)=0$,
then $h_{ab}$ is zero on $\Omega$ up to a scalar gauge transformation.
\end{prop}
In order to prove Proposition \ref{tstpos}, we write $T_{\mathrm{lg}}^{cd}$ in the form
\beq
\label{eq.lg.tss}
T_{\mathrm{lg}}^{cd} = \sum_{i=1}^3 2\left(-F^{ca}{}_i F^d{}_{ai} 
+ \frac{1}{4} \eta^{cd}F_{abi}F^{ab}{}_i\right),
\eeq
making use of $F_a=0$ and $F_{ab0}=0$.
(\ref{eq.lg.tss}) clearly shows that $T_{\mathrm{lg}}^{cd}$ satisfies the d.e.c.,
because for each value of $i$ the expression on the right hand side in brackets is formally the same
as the standard energy-momentum tensor $T^{cd}_{\mathrm{em}} = - F^{ca}F^d{}_a + \frac{1}{4}\eta^{cd}F_{ab}F^{ab}$
of the electromagnetic field,
and $T^{cd}_{\mathrm{em}}$ satisfies the d.e.c.\ 
for arbitrary antisymmetric tensor $F_{ab}$ (see, e.g., \cite{PR} for the proof of the latter statement).
Concerning the second statement, $T_{\mathrm{lg}}^{00}$ can be written as
\beq
\label{eq.tst00}
T_{\mathrm{lg}}^{00} = \sum_{i,j=1}^3 F_{0ji}F_{0ji}
+ \frac{1}{2}\sum_{i,j,k=1}^3 F_{jki}F_{jki},
\eeq
which shows that if $T_{\mathrm{lg}}^{00}=0$, then $F_{0ji}=F_{jki}=0$.
Together with $F_{ab0}=0$, this means that $F_{abc}=0$. 
According to Lemma \ref{Fabc0},
if $F_{abc}=0$ on $\Omega$ and $H_{\mathrm{dR}}^1(\Omega)=0$,
then $h_{ab}$ is zero on $\Omega$
up to a scalar gauge transformation.
\hfill$\blacksquare$\\

As $F_a=0$ and $F_{ab0}=0$ hold in TT gauge,
Proposition \ref{tstpos} implies that $T_{\mathrm{lg}}^{cd}$ satisfies the d.e.c.\ 
in TT gauge.

\subsection{Symmetric energy-momentum tensors in generalized harmonic gauges}
\label{sec.symemtgh}

In \cite{BS} it was found (see Sections VI.B.3, VI.B.4) that in generalized harmonic gauges 
there exists a $3$-parameter family of symmetric energy-momentum tensors (degenerating to a $2$-parameter family if $\chi=1$).
Our aim in this section is to better understand these tensors, making use of the Fierz tensor.
It is also explained why a larger ($5$-parameter) family of not necessarily symmetric
energy-momentum tensors was found in \cite{BS} in generalized harmonic gauges
than without gauge fixing.

The linearized Landau--Lifshitz pseudotensor is symmetric, therefore it is obviously a member of the $3$-parameter family.
As shown below, two additional distinguished members are
\bea
\label{eq.lg.Ts}
T_{\mathrm{s}}^{cd} & = & 2\left(F^{cab}F^d{}_{ab}
- \frac{1}{4} \eta^{cd} F^{eab}F_{eab}\right) \\
\label{eq.Ts1}
T_{\mathrm{s'}}^{cd} & = & T_{\mathrm{s}}^{cd}
- \left( F^c F^d - \frac{1}{2}\eta^{cd}F^a F_a\right);
\eea
$T_{\mathrm{LL}}^{cd}$, $T_{\mathrm{s}}^{cd}$ and $T_{\mathrm{s'}}^{cd}$ thus span the $3$-parameter family.

Concerning $T_{\mathrm{s}}^{cd}$,
in Section \ref{sec.tlgpos} we saw that $T_{\mathrm{lg}}^{cd} = T_{\mathrm{s}}^{cd}$ in the $F_a=0$ gauge,
therefore $T_{\mathrm{s}}^{cd}$ is conserved in this gauge.
Nevertheless, $T_{\mathrm{s}}^{cd}$ is conserved 
even in the much more general gauge defined by the condition $\partial_a F_b-\partial_bF_a=0$,
and thus also in any generalized harmonic gauge, as can be seen from the identity
\beq
\label{eq.divTs}
\partial_c T_{\mathrm{s}}^{cd} = -2G^{ab}F^d{}_{ab} -\frac{1}{2}F^{abd}(\partial_aF_b-\partial_bF_a)
+F_a(\partial^aF^d-\partial^dF^a).
\eeq
This identity also shows that $T_{\mathrm{s}}^{cd}$ is indeed a member of the $3$-parameter family found in \cite{BS},
since it has, taking into account (\ref{eq.fx}), the form required in \cite{BS}.

Similarly, for $T_{\mathrm{s'}}^{cd}$ the identity
\beq
\label{eq.divTsp}
\partial_c T_{\mathrm{s'}}^{cd} = -2G^{ab}F^d{}_{ab} + G_a{}^aF^d - \frac{1}{2}F^{abd}(\partial_aF_b-\partial_bF_a)
\eeq
holds, showing that $T_{\mathrm{s'}}^{cd}$ is also a member of the $3$-parameter family of symmetric energy-momentum tensors
in generalized harmonic gauges and is conserved even in the
$\partial_a F_b-\partial_bF_a=0$ gauge.
$T_{\mathrm{s'}}^{cd}$ is also obviously equal to $T_{\mathrm{lg}}^{cd}$ in the $F_a=0$ gauge.

It is not difficult to verify by direct calculations that
$T_{\mathrm{s}}^{cd}$ is the only traceless member (up to rescaling by a constant) of the $3$-parameter family
spanned by $T_{\mathrm{LL}}^{cd}$, $T_{\mathrm{s}}^{cd}$ and $T_{\mathrm{s'}}^{cd}$.
$T_{\mathrm{s'}}^{cd}$ is generally not traceless, except in the case $\chi=1$, when it coincides with
$T_{\mathrm{s}}^{cd}$. 
$T_{\mathrm{s}}^{cd}$ is duality symmetric as well, i.e.\ it does not change if $F_{abc}$ is replaced by $\tilde{F}_{abc}$.
For this property only the antisymmetry of $F_{abc}$ in the first two indices has significance.

$T_{\mathrm{s'}}^{cd}=T_{\mathrm{lg}}^{cd} + (1-\chi)F^{cad}\partial_a h - F^{cad}X_a(\chi)$,
and $F^{cad}\partial_a h$ is a trivially conserved tensor already encountered in Section \ref{sec.4p},
therefore $T_{\mathrm{s'}}^{cd}$ is a member of the $4$-parameter family (\ref{eq.lg.em}) up to the term
$-F^{cad}X_a(\chi)$, which is zero in generalized harmonic gauge. 
This implies that 
$T_{\mathrm{s'}}^{cd}$ gives the same total energy and momentum
as $T_{\mathrm{lg}}^{cd}$ in generalized harmonic gauges, if $h_{ab}$ and $\partial_c h_{ab}$
fall off sufficiently rapidly at infinity.

On the other hand, $T_{\mathrm{s}}^{cd}$ generally gives different total energy and momentum
than $T_{\mathrm{lg}}^{cd}$ and $T_{\mathrm{s'}}^{cd}$, since the difference
\beq
\label{eq.Ttr1}
T_{\mathrm{tr}}^{cd}=
T_{\mathrm{s}}^{cd}-T_{\mathrm{s'}}^{cd} = F^c F^d - \frac{1}{2}\eta^{cd}F^a F_a
\eeq
satisfies the d.e.c.\ and thus generally gives nonzero contribution to the total energy and momentum.
(In particular, $T_{\mathrm{tr}}^{00}=\frac{1}{2}(F_0F_0+\sum_i F_iF_i)$.
The validity of the d.e.c.\ for $T_{\mathrm{tr}}^{cd}$ can be proved in the same way as for the energy-momentum tensor of the scalar field.)
In view of this property, $T_{\mathrm{s}}^{cd}$ does not appear to be a proper energy-momentum tensor
of the linearized gravitational field outside the $F_a=0$ gauge,
in spite of its strong formal similarity to $T_{\mathrm{em}}^{cd}$.

The above property of $T_{\mathrm{s}}^{cd}$ also implies that $T_{\mathrm{s}}^{cd}$
is not a member of the $4$-parameter family (\ref{eq.lg.em}) even in generalized harmonic gauges,
with the exception of the $F_a=0$ gauge.
This is consistent with the result of \cite{BS} that in generalized harmonic gauges there exists
(under the conditions imposed in \cite{BS}) a larger family of energy-momentum tensors
than without gauge fixing. That family has $5$ free parameters, therefore it can be spanned by $T_{\mathrm{s}}^{cd}$ and (\ref{eq.lg.em}).

Concerning $T_{\mathrm{tr}}^{cd}$,
it is also worth mentioning that in generalized harmonic gauge
\beq
\label{eq.Ttr2}
T_{\mathrm{tr}}^{cd} = (\chi-1)^2 \left(\partial^c h\partial^d h - \frac{1}{2}\eta^{cd}\partial_ah\partial^ah\right).
\eeq
Up to normalization, this is just
the energy-momentum tensor of $h$ as a scalar field subject to the wave equation $\Box h=0$,
which indeed holds in generalized harmonic gauge if\\
$\chi\ne 1$.
Thus, although $T_{\mathrm{tr}}^{cd}$ is allowed by the conditions imposed on gravitational energy-momentum tensors in \cite{BS},
it is in fact an energy-momentum tensor for the trace of $h_{ab}$, rather than for $h_{ab}$ itself.

In view of the above considerations,
we can say that the following two circumstances are responsible for
the larger number of symmetric energy-momentum tensors in generalized harmonic gauges
than without gauge fixing:\\[1mm]
(a) in generalized harmonic gauge the member
$T_{\mathrm{lg}}^{cd} + (1-\chi)F^{cad}\partial_a h$
of the $4$-parameter family (\ref{eq.lg.em}) also becomes symmetric,\\
(b) $T_{\mathrm{tr}}^{cd}$ appears as a new conserved symmetric tensor in generalized harmonic gauges,
associated with the wave equation $\Box h=0$ that becomes valid in these gauges.\\[1mm]
The appearance of $T_{\mathrm{tr}}^{cd}$ also explains why a larger family of
not necessarily symmetric energy-momentum tensors was found in \cite{BS} in generalized harmonic gauges.

We note that the formulae presented in \cite{BS} for the $3$-parameter family of symmetric energy-momentum tensors
contain $\chi$ explicitly, whereas $T_{\mathrm{LL}}^{cd}$, $T_{\mathrm{s}}^{cd}$ and $T_{\mathrm{s'}}^{cd}$
do not. 
Nevertheless, the tensors given in \cite{BS}
can be rewritten in $\chi$-independent form by
replacing $\chi\partial_a h$ with $\partial_b h^b{}_a$,
making use of the generalized harmonic gauge condition.
The same is true also for the expression for the $5$-parameter family of not necessarily symmetric tensors
found in \cite{BS} (see equation (43)).

In harmonic gauge there exists a further distinguished symmetric energy-momentum tensor,
which was advocated in \cite{BHL1}; it will be discussed in Section \ref{sec.kg}.

\subsection{Klein--Gordon-type Lagrangian and energy-momentum tensor}
\label{sec.kg}

In this subsection 
we point out that the energy-momentum tensor
\beq
\label{eq.lg.TKG}
\tau^{cd} = \frac{1}{4}\left(\partial^c h^{ab} \partial^d h_{ab} - \frac{1}{2}\eta^{cd}\partial_e h_{ab}\partial^e h^{ab}\right)
-\frac{1}{8}\left(\partial^c h \partial^d h - \frac{1}{2}\eta^{cd}\partial_a h \partial^a h\right),
\eeq
proposed in \cite{BHL1}, can be reproduced as a canonical energy-momentum tensor
following from a simple gauge fixed Lagrangian. 
This way of generating $\tau^{cd}$ complements the Einstein--Hilbert-type procedure presented in \cite{BHL2}.
We also give a short derivation for the result of \cite{BHL1} that $\tau^{cd}$
satisfies the dominant energy condition in TT gauge,
and show that $\tau^{cd}$ has an analogue in electrodynamics.

Let us consider the gauge fixed Lagrangian
\beq
\label{eq.lg.LKG}
L_{\mathrm{KG}} = L_1+\frac{1}{4}L_2+L_{\mathrm{h}} 
= \frac{1}{4}\partial_a h_{bc} \partial^a h^{bc} -\frac{1}{8}\partial_a h \partial^a h,
\eeq
where $L_{\mathrm{h}} = \frac{1}{2}\partial_a\bar{h}^{ab} \partial_c\bar{h}^c{}_b$, $\chi=\frac{1}{2}$,
is the gauge fixing term.
This is a well-known Lagrangian for the linearized gravitational field;
it can be found in Section 2.2 of \cite{Maggiore}, for example.
It is clearly similar to the Lagrangian of the massless scalar field. 
The Euler--Lagrange equations following from $L_{\mathrm{KG}}$ are $\Box \bar{h}^{ab}=0$, which coincide with the 
linearized Einstein equations without matter in harmonic gauge.
The canonical energy-momentum tensor following from $L_{\mathrm{KG}}$ turns out to be $2\tau^{cd}$.

In \cite{BHL2}, $\tau^{cd}$ was generalized beyond harmonic gauge---see equation (13a).
By direct calculation one can verify that this generalized tensor is, up to a normalization factor $1/2$,
the canonical energy-momentum tensor following from $L_{\mathrm{KG}}$ without the gauge fixing term.
(The generalized $\tau^{cd}$ is not symmetric.
For comparison with \cite{BHL2}, we note that 
the order of the indices in \cite{BHL2} is the opposite of the order following from our definition
of the canonical energy-momentum tensor in Section \ref{sec.gauge}.)

In Section III.B of \cite{BHL1}, it was demonstrated that $\tau^{cd}$
satisfies the d.e.c.\ in TT gauge.
Here we give another, shorter proof of this result, 
in which the d.e.c.\ for $\tau^{cd}$
is traced to the d.e.c.\ 
for the scalar field, and only $h_{i0}=0$, $i=1,2,3$, (instead of $h_{a0}=0$) is used.
The proof is the following:
since the equalities $h=0$ and $h_{i0}=0$, $i=1,2,3$, are assumed to hold in an extended domain,
rather than only at a single point, they imply $\partial_a h=0$ and $\partial_a h_{i0}=0$.
Using the latter equations,
$\tau^{cd}$ can be written in the form
\beq
\label{eq.TKG2}
\tau^{cd}=\frac{1}{4}\sum_{i,j=1}^3 \left(\partial^c h_{ij}\partial^d h_{ij}
-\frac{1}{2}\eta^{cd}\partial_a h_{ij}\partial^a h_{ij}\right)
+\frac{1}{4}\left(\partial^c h_{00}\partial^d h_{00} - \frac{1}{2}\eta^{cd}\partial_a h_{00}\partial^a h_{00}\right).
\eeq
Formally this is the same as the energy-momentum tensor of ten scalar fields,
therefore it satisfies the d.e.c.

We note that
instead of the condition $h_{a0}=0$, one could assume $h_{ab}n^b=0$, where $n^b$ is a fixed
normalized timelike vector, or instead of $h_{i0}=0$, one could assume $e^ah_{ab}n^b=0$,
for all vectors $e^a$ orthogonal to $n^a$.
Nevertheless, considering only the case $n^a=(1,0,0,0)$
does not cause any loss of generality, 
since $n^a=(1,0,0,0)$ can always be achieved by a change of coordinates.

Since the $3$-parameter family of symmetric energy-momentum tensors in generalized harmonic gauges is spanned by
$T_{\mathrm{s}}^{cd}$, $T_{\mathrm{s'}}^{cd}$ and $T_{\mathrm{LL}}^{cd}$, it has to be possible to express
$\tau^{cd}$ as a linear combination of these tensors,
up to terms that vanish in harmonic gauge.
It is not difficult to see that the relation (valid in harmonic gauge) is
\beq
\label{eq.tauTT}
\tau^{cd} = T_{\mathrm{s'}}^{cd}-T_{\mathrm{LL}}^{cd}.
\eeq
This equation shows that in TT gauge the trace of $\tau^{cd}$ is $-1$ times the trace of $T_{\mathrm{LL}}^{cd}$.

For a direct comparison of the energy densities, we write down explicit formulae for them in TT gauge.
$T_{\mathrm{lg}}^{00}$ is
\bea
&& \frac{1}{4}\biggl(\ \sum_{i,j} (\partial_0 h_{ij})^2
+ \sum_{i,j,k} \frac{1}{2}(\partial_i h_{jk} - \partial_j h_{ik})(\partial_i h_{jk} - \partial_j h_{ik}) \biggr) \nonumber \\
&&\hspace{7.5mm} = \frac{1}{4}\biggl(\ \sum_{i,j} (\partial_0 h_{ij})^2
+ \sum_{i,j,k} (\partial_i h_{jk}\partial_i h_{jk} - \partial_i h_{jk}\partial_j h_{ik})\biggr) \nonumber
\eea
in TT gauge, 
whereas 
$\tau^{00}=
\frac{1}{8}\bigl( \sum_{i,j} (\partial_0 h_{ij})^2
+ \sum_{i,j,k} \partial_i h_{jk}\partial_i h_{jk}\bigr)$.
The canonical energy-momentum density $\Tc^{00}$ is equal to $T_{\mathrm{lg}}^{00}$.
For the linearized Landau--Lifshitz pseudotensor
$T_{\mathrm{LL}}^{cd} = T_{\mathrm{lg}}^{cd} - \tau^{cd}$ holds in TT gauge.
For a comparison of $\tau^{cd}$ with the ADM energy momentum \cite{ADM1,ADM2,ADM3}, see \cite{BHL2}.\\

%%%%%%%%%%%%%%%%%%%%%%%%%%%%%%%%%%%%%%%%%%%%%%%%%%%%%%%%%%%%%%%%%%%%%%%%%%%%%%%%%%%%%%%%%%%%%%%%%%%%%

\noi
{\bf Analogue of $\tau^{cd}$ in electrodynamics}\\
The electromagnetic field also has a well-known gauge fixed Klein--Gordon-type Lagran-gian:
\beq
L_{\mathrm{em,KG}} = L_{1,\mathrm{em}} - \frac{1}{2} L_{2,\mathrm{em}} - \frac{1}{2}(\partial_a A^a)^2 = 
-\frac{1}{2}(\partial_a A_b)(\partial^a A^b),
\eeq
where the gauge fixing term is $-\frac{1}{2}(\partial_a A^a)^2$
($L_{1,\mathrm{em}}$ and $L_{2,\mathrm{em}}$ are defined in (\ref{eq.eml2})).
The corresponding Euler--Lagrange equations are $\Box A^a = 0$, which coincide with
Maxwell's equations in Lorenz gauge. 
The canonical energy-momentum tensor following from $L_{\mathrm{em,KG}}$ is
\beq
\label{eq.TemKG}
T^{ab}_{\mathrm{em,KG}} = -\partial^a A^c \partial^b A_c +\frac{1}{2}\eta^{ab} (\partial_c A_d)(\partial^c A^d).
\eeq
Similarly to $\tau^{ab}$,
$T^{ab}_{\mathrm{em,KG}}$ is symmetric but not traceless, and it satisfies the d.e.c.\ in transverse gauge
(which is characterized by $A_0=0$, in addition to $\partial_a A^a=0$).
The latter property can be proved in the same way as in the case of $\tau^{ab}$:
in virtue of $A_0=0$, we have 
\beq 
\label{eq.emtkg2}
T^{ab}_{\mathrm{em,KG}} = \sum_{i=1}^3\left( \partial^a A_i \partial^b A_i 
- \frac{1}{2} \eta^{ab} (\partial_c A_i)(\partial^c A_i) \right),
\eeq 
where the right hand side is formally the same as the energy-momentum tensor of three real scalar fields,
thus the validity of the d.e.c.\ for $T^{ab}_{\mathrm{em,KG}}$
follows from its validity for the scalar field.

\newpage

\section{Summary of the properties of $T_{\mathrm{lg}}^{cd}$}
\label{sec.emtsum}

In this section the main results of the previous sections concerning $T_{\mathrm{lg}}^{cd}$
are collected.

\begin{enumerate}
\item 
$T_{\mathrm{lg}}^{cd}$ can be expressed in terms of $F_{abc}$ (see (\ref{eq.lg.Tst})) and is thus scalar gauge invariant.
In the form (\ref{eq.lg.Tst}) it shows considerable similarity to the usual energy-momentum tensor of the electromagnetic field,
especially if the form (\ref{eq.lg.foe}) of the gravitational field equations is taken into account.
Gauge fixing is not required for its definition.
It is traceless, but not symmetric.
It becomes symmetric if $F_a=0$,
therefore it is also symmetric in traceless harmonic gauge and in TT gauge.
It is invariant with respect to the duality $F_{abc}\to\tilde{F}_{abc}$ if $F_a=0$
(see Section \ref{sec.lg.duality} for further details on this duality transformation). 

\item
Similarly to the energy-momentum tensor of the electromagnetic field,
$T_{\mathrm{lg}}^{cd}$ can be obtained by adding a trivially conserved correction term to the canonical
energy-momentum tensor following from $L_1$
(see equations (\ref{eq.lg.l1}), (\ref{eq.lg.x7}) and (\ref{eq.L1forms}) for the definition of $L_1$).
Furthermore, for any vector $e^a$, $T_{\mathrm{lg}}^{cd}e_d$ is a Noether current associated with
the variation $\delta h_{ab} = (\mathring{F}_{cab}+\mathring{F}_{cba})e^c$,
which is scalar gauge invariant
and corresponds to an infinitesimal translation in the direction $e^a$
accompanied by a field-dependent infinitesimal gauge transformation.

\item
$T_{\mathrm{lg}}^{cd}$ changes by a trivially conserved tensor field under general gauge transformations.

\item
$T_{\mathrm{lg}}^{cd}$ satisfies the d.e.c.\ if $F_a=0$ and $F_{ab0}=0$,
therefore it also satisfies the d.e.c.\ in TT gauge.

In the $F_a=0$, $F_{ab0}=0$ gauge, if the energy density $T_{\mathrm{lg}}^{00}$ is zero 
in a spacetime domain $\Omega$ with $H^1_{\mathrm{dR}}(\Omega)=0$,
then $h_{ab}$ is zero in $\Omega$ up to a scalar gauge transformation.

\item
$T_{\mathrm{lg}}^{cd}$ is the only traceless tensor
and the only scalar gauge invariant tensor, up to normalization,
in the general $4$-parameter family (\ref{eq.lg.em}) of energy-momentum tensors found in \cite{BS}
without gauge fixing.
(There is also a single symmetric tensor in this family,
namely the linearized Landau--Lifshitz pseudotensor $T_{\mathrm{LL}}^{cd}$.)

\item
The total energy and momentum given by
$T_{\mathrm{LL}}^{cd}$, $\tau^{cd}$ and $T_{\mathrm{lg}}^{cd}$ agree,
up to a normalization factor $2$,
if $h_{ab}$ and $\partial_c h_{ab}$ fall off sufficiently rapidly at infinity.
Nevertheless, $T_{\mathrm{lg}}^{cd}$ generally differs from $2T_{\mathrm{LL}}^{cd}$ and $2\tau^{cd}$ even in TT gauge.

\item
$T_{\mathrm{lg}}^{cd}$
and the entire family (\ref{eq.lg.em}) have extensions in the presence of matter
that satisfy, up to normalization, the balance equation (\ref{eq.Tbal}), required in the framework introduced in \cite{BHL1}.

\item
The results of \cite{BHL1,BHL2} show that the harmonic gauge and the TT gauge
are closely associated with $\tau^{cd}$;
these gauges are important for the symmetry and the positivity of $\tau^{cd}$, respectively.
From the properties of $T_{\mathrm{lg}}^{cd}$ described above, it can be seen that there are similar gauges
in the case of $T_{\mathrm{lg}}^{cd}$ as well, namely the $F_a=0$ gauge and the $F_a=0$, $F_{ab0}=0$ gauge.
\end{enumerate}

The asymmetry of $T_{\mathrm{lg}}^{cd}$ may be regarded as an imperfection,
nevertheless it is in good agreement
with the asymmetry of the first order Maxwell-like field equations (\ref{eq.lg.foe}).
A similar correspondence between the form of the first order field equations and the symmetry of the
energy-momentum tensor can be seen in the case of
$T_{\mathrm{s'}}^{cd}$ (\ref{eq.Ts1}) and $T_{\mathrm{s}}^{cd}$ (\ref{eq.lg.Ts}) as well,
as the first order Maxwell-like field equations
have a symmetric form (\ref{eq.lg.foeharm}) in generalized harmonic gauges.
The asymmetry of $T_{\mathrm{lg}}^{cd}$ and the symmetry of $T_{\mathrm{s'}}^{cd}$ and $T_{\mathrm{s}}^{cd}$
are thus not accidental, but are reflections of the structure of the
first order Maxwell-like form of the field equations
of linearized gravity.

\section{Duality symmetry of the linearized gravitational field}
\label{sec.lg.duality}

In this section we show that the linearized Einstein equations without matter
have a duality symmetry in the gauge $F_a=0$,
which maps $F_{abc}$ to $\tilde{F}_{abc}$.
Since $F_{abc}$ is not completely gauge invariant, it is not immediately obvious if
this duality symmetry can be restricted to a particular subgauge of the $F_a=0$ gauge that one might wish
to use. Regarding this problem,
we show that the duality we introduce can be restricted to traceless harmonic gauge, to TT gauge
and to the $F_a=0$, $F_{ab0}=0$ gauge.
Remarkably, the case of the $F_a=0$, $F_{ab0}=0$ gauge turns out to be much simpler than the case of the TT gauge.
We also show that duality is respected by gauge transformations that preserve the $F_a=0$ gauge.
We determine the action of duality transformation on monochromatic plane waves,
and obtain,
by applying a general construction which is suitable for theories with a quadratic Lagrangian,
a conserved current associated with duality symmetry.

We define duality in the following way:
a solution $h'_{ab}$ of the linearized Einstein equations without source is a dual of a solution $h_{ab}$
if the Fierz tensor corresponding to $h'_{ab}$ is equal to $\tilde{F}_{abc}$. 
We shall use the notation $F'_{abc}$ for the Fierz tensor corresponding to $h'_{ab}$.

The dual of $h_{ab}$ is not unique;
scalar gauge transformations can obviously be applied to it, as the Fierz tensor is scalar gauge invariant. 
If $h'_{ab}$ is a dual of $h_{ab}$, then $-h_{ab}$ is clearly a dual of $h'_{ab}$.
Duality rotations can be defined in the usual way in terms of $h_{ab}$ and $h'_{ab}$.

In view of (\ref{eq.lg.dualcycl}) and the remark below it, 
if $h_{ab}$ has a dual, then $F_a=0$.
This implies, taking into account (\ref{eq.r1}), that
if $h'_{ab}$ is a dual of $h_{ab}$,
then the linearized Riemann tensor of $h'_{ab}$ is also the dual of the linearized Riemann tensor of $h_{ab}$
(i.e., it is $\tilde{R}_{abcd}$),
thus $h'_{ab}$ is also a dual of $h_{ab}$ according to the definition in \cite{HT} based on the Riemann tensor.

$F_a=0$ is not only necessary for the existence of a dual of $h_{ab}$,
but sufficient as well,
if the domain where $h_{ab}$ is defined satisfies a topological condition:   
\begin{prop}
\label{prd1}
If $h_{ab}$ satisfies the condition $F_a=0$ and 
is a solution of the linearized Einstein equations
without source on a spacetime domain $\Omega$ with $H^2_{\mathrm{dR}}(\Omega)=0$,
then there exists a dual of $h_{ab}$ on $\Omega$. 
\end{prop}
This statement can be proved in a straightforward way:
from $F_a=0$ it follows that $\tilde{F}_{abc}$ has the cyclic property (\ref{eq.lg.x2}) (see (\ref{eq.lg.dualcycl})).
Taking also into account $\partial_c F^{cab}= -\partial_c \tilde{\tilde{F}}^{cab} = 0$ and $H^2_{\mathrm{dR}}(\Omega)=0$, 
there exists, according to Lemma \ref{pr1}, a symmetric tensor field $h'_{ab}$ defined on $\Omega$ so that
$\tilde{F}_{abc} = \frac{1}{2}(\partial_a h'_{bc} - \partial_b h'_{ac})$.
$\tilde{F}_{abc}$ is traceless (see (\ref{eq.lg.x2b})), therefore $\tilde{F}_{abc}$
is the Fierz tensor corresponding to $h'_{ab}$ (see (\ref{eq.trx1}), (\ref{eq.circ2}), (\ref{eq.circx})).
From $F_a=0$ it also follows that $F_{abc}=\mathring{F}_{abc}$,
thus from $\partial_c \tilde{\mathring{F}}^{cab}=0$ (see (\ref{eq.circ3})) it follows that $\partial_c \tilde{F}^{cab}=0$,
i.e.\ $h'_{ab}$ is a solution of the linearized Einstein equations without source.
This means that $h'_{ab}$ is a dual of $h_{ab}$. \hfill$\blacksquare$\\

Proposition \ref{prd1} is the main statement concerning the existence of the duality symmetry based on the Fierz tensor. 
The condition $H^2_{\mathrm{dR}}(\Omega)=0$ is satisfied, in particular, if $\Omega$ is the entire Minkowski spacetime.\\

Since $F_{abc}$ is not invariant under all gauge transformations that preserve the $F_a=0$ gauge,
it is natural to ask if duality is respected by gauge transformations in some sense.
For answering this question, the following fact is useful: 
\begin{lemma}
\label{dginv}
If $h_{ab}$ is defined on a spacetime domain $\Omega$ with $H_{\mathrm{dR}}^2(\Omega)=0$
and is zero up to a gauge transformation that preserves the $F_a=0$ gauge,
then it has a dual on $\Omega$ with the same property.
\end{lemma}
In order to prove this statement, we first note that
$h_{ab}$ can be written, according to the assumption,
in the form $h_{ab}=\partial_a\phi_b+\partial_b\phi_a$.
For field configurations of this form,
equation $F_a=0$ coincides with Maxwell's equations for $\phi_a$,
i.e.\ with the equation $\partial_a f^{ab}=0$, where $f_{ab}=\partial_a\phi_b-\partial_b\phi_a$.
From $\partial_a f^{ab}=0$ it follows that
$F_{abc}=\frac{1}{2}\partial_c f_{ab}$, therefore
$\tilde{F}_{abc}=\frac{1}{2}\partial_c\tilde{f}_{ab}$.
In addition, $H_{\mathrm{dR}}^2(\Omega)=0$ and $\partial_a f^{ab}=0$ imply that there exists
a covector field $\phi'_a$ on $\Omega$ so that
$\tilde{f}_{ab}=\partial_a\phi'_b-\partial_b\phi'_a$.
From $f_{ab}=\partial_a\phi_b-\partial_b\phi_a$ it follows that $\partial_a\tilde{f}^{ab}=0$,
i.e.\ $\phi'_a$ satisfies Maxwell's equations.
$h'_{ab}=\partial_a\phi'_b+\partial_b\phi'_a$ is thus a symmetric tensor field
that is zero up to a gauge transformation
that preserves the $F_a=0$ gauge (the parameter of the gauge transformation being $\phi'_a$).
The Fierz tensor of $h'_{ab}$
is $\frac{1}{2}\partial_c\tilde{f}_{ab}$, and $\frac{1}{2}\partial_c\tilde{f}_{ab} = \tilde{F}_{abc}$,
as we have seen, therefore $h'_{ab}$ is a dual of $h_{ab}$.
\hfill$\blacksquare$\\

Using Lemma \ref{dginv}, it is easy to see that gauge transformations respect duality in the following sense:
\begin{prop}
If $h^{(1)}_{ab}$ and $h^{(2)}_{ab}$, defined on 
a spacetime domain $\Omega$ with $H_{\mathrm{dR}}^2(\Omega)=0$,
are solutions of the linearized Einstein equations without source in the $F_a=0$ gauge
and are equal up to a gauge transformation,
and $k^{(1)}_{ab}$ is a dual of $h^{(1)}_{ab}$ on $\Omega$,
then $h^{(2)}_{ab}$ has a dual on $\Omega$ that is equal to $k^{(1)}_{ab}$
up to a gauge transformation.
\end{prop}

\subsection{Restriction to subgauges of the $F_a=0$ gauge}

Regarding the restriction of the duality introduced above to subgauges of the $F_a=0$ gauge,
the case of the gauge $F_a=0$, $F_{ab0}=0$ is very simple:
it is obvious that if $F_{ab0}=0$, then $\tilde{F}_{ab0}=0$,
i.e.\ the gauge $F_a=0$, $F_{ab0}=0$ is preserved by duality.

For the traceless harmonic gauge, the following statement can be made:
\begin{prop}
\label{prd2}
If $h_{ab}$ satisfies the traceless harmonic gauge conditions and 
is a solution of the linearized Einstein equations
without source on a spacetime domain $\Omega$ with $H^2_{\mathrm{dR}}(\Omega)=0$,
and $U$ is a bounded domain with the property $\overline{U}\subset \Omega$,
then $h_{ab}$ has a dual, defined on $U$,
that satisfies the traceless harmonic gauge conditions.
\end{prop}
From Proposition \ref{prd1} it follows that
there exists a dual $h'_{ab}$ of $h_{ab}$ defined on $\Omega$,
therefore in order to prove Proposition \ref{prd2}
it is sufficient to show that it is possible to choose $h'_{ab}$ so that its trace is zero.
To do this, one can subject $h'_{ab}$ to a scalar gauge transformation,
i.e.\ replace $h'_{ab}$ with $h''_{ab}=h'_{ab} + \partial_{ab}\phi$,
and try to choose the parameter function $\phi$
so that $h''_a{}^a=0$.
$h''_a{}^a=0$ is achieved if
$\phi$ satisfies the inhomogeneous wave equation $\Box \phi = - h'_a{}^a$, i.e.\ one should find a solution to this equation.
This can be done by the Green's function method on any bounded domain
$U$ that has the property $\overline{U}\subset \Omega$.
In the convolution one can use an extension of the function $-h'_a{}^a|_U$ to a bump function on a slightly larger set.
Such an extension of $-h'_a{}^a|_U$ is, for instance, $-h'_a{}^a B$, where $B$ is a bump function 
with the properties $B|_U=1$ and $\mathrm{supp}(B)\subset\Omega$.
\hfill$\blacksquare$\\

$U$ is included in Proposition \ref{prd2}
to ensure that the inhomogeneous wave equation $\Box \phi = - h'_a{}^a$ can be solved straightforwardly using
the Green's function method. 
Nevertheless, suitable solutions exist under different conditions as well---for instance,
if $\Omega$ is convex, then $U$ can be omitted altogether,
as on convex domains the wave equation has a smooth solution with arbitrary smooth source \cite{Hormander}.
This means, in particular, that if $h_{ab}$ is defined on the entire Minkowski spacetime,
then it has a dual (in traceless harmonic gauge) that is also defined on the entire Minkowski spacetime.
Further assumptions about $h_{ab}$, e.g.\ fall off conditions at infinity, are not needed for this.\\

In the case of the TT gauge, 
the statement about the existence of duality symmetry is modified in the following way:
\begin{prop}
\label{prd3}
Let $h_{ab}$ be a symmetric tensor field that satisfies the TT gauge conditions and 
the linearized Einstein equations
without source on a spacetime domain $\Omega$ with $H^2_{\mathrm{dR}}(\Omega)=0$.
Let $U$ be a bounded domain with the properties
$\overline{U}\subset \Omega$ and $H^1_{\mathrm{dR}}(U)=0$.
Let $\mc{R}$ be a region of the form $[t_0,t_1]\times \Sigma$, where $\Sigma$ is a three-dimensional spatial domain,
so that $\overline{\mc{R}} \subset U$.
Then $h_{ab}$ has a dual, defined on $\mc{R}$,
that satisfies the TT gauge conditions.
\end{prop}
The proof is the following: according to Proposition \ref{prd2} about the duality symmetry in
traceless harmonic gauge, there exists an $h'_{ab}$, defined on $U$, that has all the required properties except $h'_{a0}=0$.
$h'_{ab}$ can be modified freely by a scalar gauge transformation $h'_{ab} \to h''_{ab}=h'_{ab}+\partial_{ab}\phi$,
where $\phi$ is a function with the property $\Box \phi=0$.
The traceless harmonic gauge conditions are satisfied by 
$h''_{ab}$ because $h'_{ab}$ satisfies them and $\Box \phi=0$.
Using this gauge freedom, we try to choose $\phi$ so that $h''_{a0}=0$.

From $F_{ab0}=0$ it follows that $\tilde{F}_{ab0}=0$, which implies $\partial_a h'_{b0}-\partial_b h'_{a0}=0$.
From the latter equation it follows that
$h'_{a0}=\partial_a q$ with some function $q$, as $H^1_{\mathrm{dR}}(U)=0$.
From the traceless harmonic gauge conditions it follows that $\partial_a h'^a{}_{0}=\Box q = 0$. 
$h''_{a0}=0$ is equivalent with $\partial_a\partial_0 \phi = - h'_{a0} = - \partial_a q$,
therefore $h''_{a0}=0$ is achieved if we can find a $\phi$ so that $\partial_0 \phi = -q$. 
A $\phi$ with this property, also satisfying the condition $\Box \phi=0$,
can be constructed on $\mc{R}$ by choosing $\phi(t_0,\mr{x})$ (where $\mr{x}\in\Sigma$)
so that $\partial_i\partial^i \phi (t_0,\mr{x}) = \partial_0 q(t_0,\mr{x})$
and then defining $\phi(t,\mr{x})$ as $\phi(t,\mr{x})=\phi(t_0,\mr{x}) + \int_{t_0}^{t} -q(\tau,\mr{x})  d\tau$, $t\in [t_0,t_1]$.
It is easy to verify by direct calculations that the $\phi$ constructed in this way has the properties
$\Box \phi=0$, $\partial_0 \phi = -q$.
Finding $\phi(t_0,\mr{x})$ involves solving the Poisson equation
$\partial_i\partial^i \phi (t_0,\mr{x}) = \partial_0 q (t_0,\mr{x})$ on $\Sigma$;
this can be done by applying the Green's function method.
Similarly as in the proof of Proposition \ref{prd2}, $\partial_0 q(t_0,\mr{x})$ can be extended from $\Sigma$
to a bump function on a slightly larger three-dimensional surface containing $\Sigma$
($\Sigma$ is bounded because of the boundedness of $U$),
and one can use this function in the convolution with the usual Green's function of the Laplace operator.
\hfill$\blacksquare$\\

As in the case of the traceless harmonic gauge, $U$ is included in Proposition \ref{prd3}
to ensure that the Green's function method can be used straightforwardly to solve the inhomogeneous wave equation and
the Poisson equation. Nevertheless, these equations have solutions under different conditions as well; in particular if $\Omega$
is the entire Minkowski spacetime, then
$U$ can be omitted completely, $\mc{R}=\RR\times \RR^3$ can be chosen,
and $\phi$ can be defined as $\phi(t,\mr{x})=\phi(0,\mr{x}) + \int_{0}^{t} -q(\tau,\mr{x})  d\tau$, $t\in \RR $,
because the inhomogeneous wave equation and the Poisson equation have smooth solutions on $\RR^4$ and $\RR^3$, respectively,
with arbitrary smooth source \cite{Hormander}.
Thus if $\Omega$ is the entire Minkowski spacetime, then $h_{ab}$ has a dual in TT gauge that
is also defined on the entire Minkowski spacetime.
Again, it should be emphasized that this conclusion holds without any restriction on
the behaviour of $h_{ab}$ at infinity. \\

The action of duality on $E_{ij}$ and $B_{ij}$ in the $F_a=0$, $F_{ab0}=0$ gauge and in TT gauge is
\beq
\label{eq.ebdual}
E_{ij}\to B_{ij},\qquad B_{ij} \to -E_{ij},
\eeq
as can be seen from (\ref{eq.eij}).
The Maxwell-like equations (\ref{max1})-(\ref{max3})
are clearly symmetric under this transformation. 
(\ref{eq.ebdual}) is also in agreement with \cite{Barnett}.

\subsection{Conserved current associated with duality symmetry}
\label{sec.helc}

In any theory that has a Lagrangian
that is homogeneous quadratic in the basic dynamical fields and their derivatives,
it is possible to associate a conserved current with
any symmetry of the equations of motion.
Depending on the symmetry, the corresponding current can be nonlocal. 
In the case of free fields, for instance, there exist conserved currents associated
with space and time reflections \cite{LTU}.
There are several different ways of constructing these currents; 
here we present an approach
that is inspired by the duality symmetric formulation of linearized gravity introduced in \cite{Barnett}
(for electrodynamics, see \cite{AncoThe, CamBar, BBN, CameronPhd, AAR}).
For a different approach, see Section 2 of \cite{TG1} and references therein.
Although here we focus on linearized gravity and duality symmetry,
we briefly describe the construction in general form as well at the end of this subsection.

Let us consider two independent copies, $h_{ab}$ and $k_{ab}$, of the linearized gravitational field,
and let us take
\beq
L_{\mathrm{d}}[h_{ab},k_{ab}] = L_1[h_{ab}] + L_1[k_{ab}]
\eeq
as their Lagrangian.
The Euler--Lagrange equations following from $L_{\mathrm{d}}$
are clearly two copies of the linearized Einstein equations, for $h_{ab}$ and $k_{ab}$.
Since $L_1$ is a homogeneous quadratic polynomial in $\partial_c h_{ab}$, $L_{\mathrm{d}}$ is invariant under the rotations
\beq
h_{ab}\to h_{ab}\cos\theta+k_{ab}\sin\theta,\qquad k_{ab}\to k_{ab}\cos\theta - h_{ab}\sin\theta .
\eeq
The corresponding Noether current (with $K^c=0$) is
\beq
\label{eq.jd}
J_{\mathrm{b}}^c = \frac{\partial L_{\mathrm{d}}}{\partial(\partial_c h_{ab})}\delta h_{ab} +
\frac{\partial L_{\mathrm{d}}}{\partial(\partial_c k_{ab})}\delta k_{ab}
= F^{cab}k_{ab} - G^{cab}h_{ab},
\eeq
where $G^{cab}$ denotes the Fierz tensor for $k_{ab}$, $\delta h_{ab}=k_{ab}$, $\delta k_{ab}=-h_{ab}$.
$J_{\mathrm{b}}^c$ is conserved if $h_{ab}$ and $k_{ab}$
are both solutions of the linearized Einstein equations, regardless of any gauge fixing.
If $k_{ab}$ is taken to be a dual of $h_{ab}$, in particular, then the current associated
with duality symmetry is obtained:
\beq
\label{eq.jd2}
J_{\mathrm{d}}^c = F^{cab}h'_{ab} - {F'}^{cab}h_{ab}.
\eeq
In TT gauge
\beq
\label{eq.jdTT}
J_{\mathrm{d}}^0 = \frac{1}{2}(B^{ij}h_{ij} - E^{ij}h'_{ij}),\qquad
J_{\mathrm{d}}^i = \frac{1}{2}(\epsilon^{ijl}B_l{}^k h'_{kj} + \epsilon^{ijl}E_l{}^k h_{kj}).
\eeq
$J_{\mathrm{d}}^0$ agrees, 
up to normalization, with the gravitational helicity density given in equation (47) of \cite{Barnett}.
$J_{\mathrm{d}}^0$ and $J_{\mathrm{d}}^i$ are also in agreement with the result of \cite{AAB} (see equations (3.2a), (3.2b)).
$J_{\mathrm{d}}^c$ thus
extends the definition of gravitational helicity given in \cite{Barnett,AAB} beyond TT gauge.
It is Lorentz covariant, depends on the derivatives of $h_{ab}$ and $h'_{ab}$
only through the respective Fierz tensors,
and it does not depend on higher than first derivatives of $h_{ab}$ and $h'_{ab}$.
(\ref{eq.jdTT}) clearly remains valid also in the $F_a=0$, $F_{ab0}=0$ gauge.

It is important to note that $J_{\mathrm{d}}^c$ is also obviously a conserved current
associated with duality symmetry
if duality is defined as in \cite{HT} (see Section I.A.),
i.e.\ as a transformation mapping the linearized Riemann tensor into its dual.

Generally, if $O$ is a symmetry operator of the linearized Einstein equations, then the associated conserved current is
$J_{\mathrm{b}}^c|_{k_{ab}=Oh_{ab}}$. $O$ can be, for instance,
an arbitrary partial differentiation of any order, or a space or time reflection.

Applying the same construction to the electromagnetic field with its usual Lagrangian gives the current
\beq
J_{\mathrm{b,em}}^c = G^{ca}A_a  - F^{ca}C_a,
\eeq
where $C_a$ denotes the second copy of the vector potential and $G_{ab}=\partial_aC_b-\partial_bC_a$
is the associated field strength tensor.
The helicity current of the electromagnetic field is then
$J_{\mathrm{b,em}}^c|_{C_a=A_a'}$, where $A_a'$ denotes a dual of $A_a$.
The latter current can be found in several articles; see \cite{AfSt,CamBar,BBN} and references therein.

Although $J_{\mathrm{b}}^c$ and $J_{\mathrm{b,em}}^c$ are not gauge invariant,
they are gauge invariant up to a trivially conserved current,
even if the gauge transformations acting on $h_{ab}$ and $k_{ab}$ or on $A_a$ and $C_a$ are different.
This gauge invariance can be proved in a very similarly way as for $T_{\mathrm{lg}}^{cd}$ (see Section \ref{sec.ginvemt}).

The generalization of the above construction to other (not necessarily special relativistic)
theories with a quadratic Lagrangian is straightforward:
If $L(\Phi_\alpha,\partial_a\Phi_\alpha,x^a)$ is a Lagrangian that is homogeneous quadratic in $\Phi_\alpha$ and $\partial_a\Phi_\alpha$,
then $L_{\mathrm{d}}[\Phi_\alpha,\Upsilon_\alpha]=L[\Phi_\alpha] + L[\Upsilon_\alpha]$
can be taken to be the Lagrangian for the doubled theory,
$\Upsilon_\alpha$ being the second copy of the array of basic dynamical fields.
Since $L$ is quadratic, $L_{\mathrm{d}}$ is invariant under the rotations
$\Phi_\alpha\to \Phi_\alpha \cos\theta + \Upsilon_\alpha\sin\theta,
\Upsilon_\alpha \to \Upsilon_\alpha \cos\theta - \Phi_\alpha \sin\theta$,
which have the infinitesimal generator
$\delta\Phi_\alpha = \Upsilon_\alpha$, $\delta\Upsilon_\alpha = -\Phi_\alpha$.
The Noether current associated with this symmetry of $L_{\mathrm{d}}[\Phi_\alpha,\Upsilon_\alpha]$
(with $K^c=0$) is
\beq
J_{\mathrm{b}}^c =
\frac{\partial L_{\mathrm{d}}}{\partial(\partial_c \Phi_\alpha)}\Upsilon_\alpha
- \frac{\partial L_{\mathrm{d}}}{\partial(\partial_c \Upsilon_\alpha)}\Phi_\alpha
=
\frac{\partial L}{\partial(\partial_c \Phi_\alpha)}\Upsilon_\alpha
-\left.\frac{\partial L}{\partial(\partial_c \Phi_\alpha)}\right|_{\Phi_\alpha\to\Upsilon_\alpha} \Phi_\alpha,
\eeq
which is conserved if $\Phi_\alpha$ and $\Upsilon_\alpha$ are both solutions of the Euler--Lagrange equations following from $L$.
If $O$ is a symmetry operator of these Euler--Lagrange equations, then
$J_{\mathrm{b}}^c[\Phi_\alpha, O\Phi_\alpha]$ is the conserved current associated with $O$.
It is also straightforward to extend this construction to Lagrangians that depend on higher than first derivatives of the fields.

\subsection{Action of duality on monochromatic plane waves}
\label{sec.plw}

For an example, let us determine the action of duality on monochromatic plane waves
$W_{ab}=w_{ab}\cos(k_c x^c + \varphi)$, where the wave vector $k_c$ is null.
Both in the $F_a=0$ gauge and in traceless harmonic gauge,
the polarization $w_{ab}$ of 
a monochromatic plane wave solution $W_{ab}$ of the linearized Einstein equations
is a linear combination of five independent polarizations:
\beq
w^{(1)}_{ab}=k_a k_b,\qquad w^{(2)}_{ab}=e_{1a}k_b+e_{1b}k_a,\qquad
w^{(3)}_{ab}=e_{2a}k_b+e_{2b}k_a,
\eeq
\beq
w^{(4)}_{ab}=e_{1a}e_{1b}-e_{2a}e_{2b},\qquad
w^{(5)}_{ab}=e_{1a}e_{2b}+e_{2a}e_{1b},
\eeq
where $e_1^a$ and $e_2^a$ are spacelike vectors
orthogonal to $k_a$ and to each other, and
$e_{1a}e_1^a = e_{2a}e_2^a = -1$.
In particular, we choose $k^a=(1,0,0,1)$ and $e_1^a=(0,1,0,0)$, $e_2^a=(0,0,1,0)$.

$W^{(4)}_{ab}$ and $W^{(5)}_{ab}$ (which have polarizations $w^{(4)}_{ab}$, $w^{(5)}_{ab}$)
are the two standard gravitational plane wave modes (known as plus- and cross-polarized modes) that satisfy
the TT gauge conditions. 
$W^{(2)}_{ab}$ and $W^{(3)}_{ab}$ are pure gauge modes, i.e.\ they have the form 
$\partial_a\phi_b+\partial_b\phi_a$: 
$W^{(2)}_{ab}=\partial_a(e_{1b}\sin(k_cx^c + \varphi)) + \partial_b(e_{1a}\sin(k_cx^c + \varphi))$,\\
$W^{(3)}_{ab}=\partial_a(e_{2b}\sin(k_cx^c + \varphi)) + \partial_b(e_{2a}\sin(k_cx^c + \varphi))$.
$W^{(1)}_{ab}$ is a pure scalar gauge mode, i.e.\ $W^{(1)}_{ab}$ is of the form $\partial_{ab}\phi$:
$W^{(1)}_{ab}=-\partial_{ab}\cos(k_cx^c + \varphi)$.
The $F_a=0$, $F_{ab0}=0$ gauge contains $W^{(1)}_{ab}$ in addition to $W^{(4)}_{ab}$ and $W^{(5)}_{ab}$.

By calculating the Fierz tensor and its dual for $W^{(1)}_{ab},\dots,W^{(5)}_{ab}$,
it can be seen that the effect of duality transformation on $W^{(2)}_{ab},\dots,W^{(5)}_{ab}$ is
\beq
W^{(4)}_{ab}\to W^{(5)}_{ab},\ \
W^{(5)}_{ab}\to -W^{(4)}_{ab},\qqquad
W^{(2)}_{ab}\to W^{(3)}_{ab},\ \ W^{(3)}_{ab}\to -W^{(2)}_{ab}.
\eeq
The Fierz tensor corresponding to a gravitational field that is a pure scalar gauge
is zero, therefore the dual of such a field can be taken to be zero or, equivalently, itself.
The dual of $W^{(1)}_{ab}$ can thus be taken to be itself.

The linearized Riemann tensor of $W^{(3)}_{ab}$, $W^{(2)}_{ab}$ and $W^{(1)}_{ab}$ is zero, as these are pure gauge modes,
therefore only the nongauge modes $W^{(4)}_{ab}$ and $W^{(5)}_{ab}$ transform nontrivially
under the duality based on the linearized Riemann tensor.

\section{Concluding remarks}
\label{sec.concl}

We have explored fundamental properties of the linearized gravitational field related to spacetime translation symmetry,
duality and gauge symmetry,
applying the Fierz formalism and the correspondence it provides between linearized gravity and electrodynamics.
We were led by this correspondence to identify a unique gravitational counterpart,
$T_{\mathrm{lg}}^{cd}$ (\ref{eq.lg.Tst}), of the standard energy-momentum tensor of the electromagnetic field.
We consider the finding of $T_{\mathrm{lg}}^{cd}$ to be the main result of the paper.
$T_{\mathrm{lg}}^{cd}$ appears, on account of its properties (see Section \ref{sec.emtsum} for a comprehensive list),
to be a very favourable energy-momentum tensor for the linearized gravitational field.
The analogy between $T_{\mathrm{lg}}^{cd}$ and the electromagnetic energy-momentum tensor is not perfect;
$T_{\mathrm{lg}}^{cd}$ is not completely gauge invariant, in particular,
and gauge fixing is needed for some of its important properties.
Nevertheless, closer analogy does not seem possible.

Among the features of $T_{\mathrm{lg}}^{cd}$,
we emphasize its tracelessness and that it satisfies the dominant energy condition in the gauge
$F_a=0$, $F_{ab0}=0$,
which is more general than the TT gauge and fits $T_{\mathrm{lg}}^{cd}$ more naturally.
The latter property of $T_{\mathrm{lg}}^{cd}$
is important because it means that
under appropriate gauge fixing conditions
the energy of the linearized gravitational field, as given by $T_{\mathrm{lg}}^{cd}$,
has physically sensible behaviour,
i.e.\ its density is positive and it does not flow faster than light.
The tracelessness of $T_{\mathrm{lg}}^{cd}$ is remarkable because
it is in accordance with the masslessness of linearized gravity \cite{ConfFT}.
In addition, $T_{\mathrm{lg}}^{cd}$ is partially gauge invariant,
in agreement with the partial gauge invariance of $F_{abc}$,
and it becomes symmetric and duality invariant if $F_a=0$.
The partial gauge invariance of $T_{\mathrm{lg}}^{cd}$ can be appreciated in view of the fact that
the linearized gravitational field does not have a completely gauge invariant energy-momentum tensor.
The tracelessness and scalar gauge invariance make $T_{\mathrm{lg}}^{cd}$ unique in the sense that
it is the only member of the general family of energy-momentum tensors (\ref{eq.lg.em}),
found in \cite{BS}, with these properties.
In this respect $T_{\mathrm{lg}}^{cd}$ is complementary to the linearized Landau--Lifshitz pseudotensor ($T_{\mathrm{LL}}^{cd}$),
which is the only symmetric tensor in the family (\ref{eq.lg.em}).
However, the linearized Landau--Lifshitz pseudotensor does not become traceless even in TT gauge.

Regarding the nature of the energy-momentum tensor $\tau^{cd}$ proposed in \cite{BHL1},
we pointed out that $\tau^{cd}$ can be obtained as a canonical energy-momentum tensor following from
a simple gauge fixed Lagrangian (namely, (\ref{eq.lg.LKG})) that is similar to the Lagrangian of the scalar field,
and that $\tau^{cd}$ has a counterpart in electrodynamics (see (\ref{eq.TemKG})).
Thus, although $\tau^{cd}$ does not appear to be special from the point of view of the Fierz formalism,
it belongs to a remarkable type, which is not specific to linearized gravity.
While $\tau^{cd}$  is similar to $T_{\mathrm{lg}}^{cd}$ in that it satisfies the dominant energy condition in TT gauge,
its trace is generally not zero.

We also found the Fierz formalism useful for determining conserved currents associated with the gauge symmetry
of linearized gravity. The currents we obtained exhibit good analogy with the corresponding currents in electrodynamics,
and could be used in giving a detailed interpretation of the  
$4$-parameter family of energy-momentum tensors (\ref{eq.lg.em}) found in \cite{BS}.

The Fierz tensor proved useful, furthermore, for better understanding the 
$3$-parameter family of symmetric energy-momentum tensors found in \cite{BS} in generalized harmonic gauges.
In particular, we identified two distinguished members, besides $T_{\mathrm{LL}}^{cd}$, of this family,
for which compact and illuminating expressions (see (\ref{eq.lg.Ts}), (\ref{eq.Ts1}))
could be written in terms of the Fierz tensor.
Concerning the physical interpretation of the members of this $3$-parameter family,
we found that it contains a $2$-parameter family of proper energy-momentum tensors,
but it also has a member (see (\ref{eq.Ttr1})) that is apparently
an energy-momentum tensor for the trace of $h_{ab}$, rather than for $h_{ab}$ itself.

The duality symmetry that we introduced
is a refinement of the well-known duality based on the linearized Riemann tensor,
but it requires gauge fixing.
It is a generalization of the duality introduced in \cite{Barnett} in TT gauge.
We constructed an associated conserved helicity current $J_{\mathrm{d}}^c$ (see (\ref{eq.jd2})),
which is Lorentz covariant, generalizes the helicity proposed in \cite{Barnett,AAB} in TT gauge to fields
in the $F_a=0$ gauge (which contains the TT gauge),
and shows good analogy with the well-known helicity current of the electromagnetic field.
It is important to note that $J_{\mathrm{d}}^c$ is also a conserved current associated with
the duality based on the Riemann tensor, and as such it does not require gauge fixing.

Another aim of this paper was to provide a detailed account of the Fierz formalism
and to further develop it.
An important part of this is the discussion of the
first order form of the field equations, in analogy with the first order Maxwell equations.
A remarkable feature of these equations (see (\ref{eq.lg.foe})), in comparison with electrodynamics,
is the asymmetry between the inhomogeneous and the homogeneous equations.
The same kind of asymmetry can be seen 
in the expression (\ref{eq.lg.Tst}) for $T_{\mathrm{lg}}^{cd}$
and in the Lagrangian $L_1$ (see (\ref{eq.L1forms})) as well.
This asymmetry disappears in the $F_a=0$ gauge, 
moreover we found that the first order field equations have a symmetric form (see (\ref{eq.lg.foeharm}))
even in the much more general gauge $\partial_a F_b-\partial_bF_a=0$,
which contains all generalized harmonic gauges.
In TT gauge, the equations (\ref{eq.lg.foe})---which are Lorentz covariant
and do not need gauge fixing---become essentially coincident with
the Maxwellian equations presented in \cite{Barnett}.

We believe that our paper demonstrates the usefulness of the perspective provided by the Fierz formalism for
studying the linearized gravitational field.
It would be interesting to see if this formalism and the results of the paper
can be extended in some way to full nonlinear general relativity or to other backgrounds.
It would also be interesting to apply $T_{\mathrm{lg}}^{cd}$,
the helicity current and the currents associated with gauge transformations
in physical problems involving weak gravitational fields.
The framework of the Fierz formalism could be used to study further symmetries of linearized gravity as well,
and it would be interesting to see if results similar to those obtained in \cite{BHL2} for $\tau^{cd}$
can be obtained also for $T_{\mathrm{lg}}^{cd}$.

% \section*{Acknowledgments}

\section*{Appendix}

\appendix

\renewcommand{\theequation}{\Alph{section}.\arabic{equation}} 
\setcounter{equation}{0}

\section{Algebraic decomposition of $F_{abc}$}
\label{app.dec}

In the following the decomposition of $F_{abc}$ into traceless and trace parts is discussed,
along with related algebraic identities. Most of the formulae in this appendix are
consequences of (\ref{eq.lg.x1}) and (\ref{eq.lg.x2})
and are independent of how $F_{abc}$ is related to $h_{ab}$.
The decomposition of tensors having the properties (\ref{eq.lg.x1}), (\ref{eq.lg.x2}) is discussed in \cite{JWK} as well.

Let us define $\check{F}_{abc}$ and $\breve{F}_{abc}$ as
\bea
\label{eq.trx1}
\check{F}_{abc} & = & \frac{1}{3}(F_a\eta_{bc} - F_b\eta_{ac})\, ,\\
\label{eq.trx2}
\breve{F}_{abc} & = & F_{abc}-\check{F}_{abc}\, .
\eea
$\check{F}_{abc}$ and $\breve{F}_{abc}$ have the same algebraic symmetry properties as $F_{abc}$ 
(i.e.\ (\ref{eq.lg.x1}), (\ref{eq.lg.x2})), furthermore
\beq
\check{F}_a = F_a,\qquad \breve{F}_a = 0\, ,
\eeq
therefore one can call $\check{F}_{abc}$ the trace part and $\breve{F}_{abc}$ the traceless part of $F_{abc}$.
These definitions are clearly suitable for any tensor having the properties (\ref{eq.lg.x1}), (\ref{eq.lg.x2}).
Applying them to $\check{F}_{abc}$ or $\breve{F}_{abc}$, one finds that 
the trace part of $\check{F}_{abc}$ is itself and the traceless part of $\breve{F}_{abc}$ is also itself.
$\check{F}_{abc}$ and $\breve{F}_{abc}$ satisfy the identities
\bea
\label{eq.xort1}
&& \check{F}_{abc}\breve{F}^{abc} = \check{F}_{abc}\breve{F}^{acb} = 0 \\
\label{eq.xort2}
&& \check{F}_{abc}\check{F}^{abc}=\frac{2}{3}F_a F^a \\
\label{eq.xort3}
&& \epsilon^{abcd}\check{F}_{abe}\check{F}_{cd}{}^e = \epsilon^{abcd}\check{F}_{abe}\breve{F}_{cd}{}^e = 0 \\
\label{eq.xort4}
&& \epsilon^{abcd}\check{F}_{abe}\check{F}_c{}^e{}_d = \epsilon^{abcd}\check{F}_{abe}\breve{F}_c{}^e{}_d = 0\, , 
\eea
which follow from the algebraic symmetry properties (\ref{eq.lg.x1}), (\ref{eq.lg.x2}) of $F_{abc}$
and from (\ref{eq.trx1}), (\ref{eq.trx2}).

Regarding $\mathring{F}_{abc}$, defined by (\ref{eq.circx}), we note that
\beq
\label{eq.circ2}
\mathring{F}_{abc}
=F_{abc}-\frac{3}{2}\check{F}_{abc} = -\frac{1}{2}(\Gamma_{abc}-\Gamma_{bac}).
\eeq
From (\ref{eq.circ2}) it can be seen that $\mathring{F}_{abc}$ can be expressed in terms of $F_{abc}$. 
(\ref{eq.circ2}) also implies $\mathring{F}_a = -\frac{1}{2}F_a$,
thus $F_{abc}$ can be expressed in terms of $\mathring{F}_{abc}$ as
\beq
\label{eq.circ4}
F_{abc} = \mathring{F}_{abc} - (\mathring{F}_a\eta_{bc} - \mathring{F}_b\eta_{ac})\, .
\eeq
(\ref{eq.circ2}) and (\ref{eq.circ4})
can be regarded as definitions of $F_{abc}$ and $\mathring{F}_{abc}$ in terms of one another.

Generally, it is clear that $F_{abc}$ can be expressed algebraically in terms of the tensor
\beq
{}^\omega F_{abc}= F_{abc}+\omega \check{F}_{abc}
\eeq
for any number $\omega\ne -1$. Since ${}^\omega F_a=(1+\omega) F_a$, the explicit formula is
\beq
F_{abc}= {}^\omega F_{abc}-\frac{\omega}{1+\omega}\, {}^\omega\check{F}_{abc}\, .
\eeq

Using the well-known properties of the finite dimensional representations of the Lorentz group,
it is not difficult to see that the linear space of rank $3$ tensors having the properties
(\ref{eq.lg.x1}) and (\ref{eq.lg.x2}) carries the 
$\bigl(\bigl(\frac{3}{2},\frac{1}{2}\bigr) + \bigl(\frac{1}{2},\frac{3}{2}\bigr)\bigr) + \bigl(\frac{1}{2},\frac{1}{2}\bigr)$
representation. $\bigl(\bigl(\frac{3}{2},\frac{1}{2}\bigr) + \bigl(\frac{1}{2},\frac{3}{2}\bigr)\bigr)$
is $16$-dimensional and contains the tensors whose trace part is zero,
whereas  
$\bigl(\frac{1}{2},\frac{1}{2}\bigr)$ is $4$-dimensional (it is the vector representation) and
contains the tensors whose traceless part is zero.

\section{Formulae for the linearized Riemann tensor and related tensors in terms of $F_{abc}$}
\label{app.riem}

In this appendix
expressions of the linearized Riemann tensor, Ricci tensor, Ricci scalar and Weyl tensor in terms of $F_{abc}$ are presented.

The linearized Christoffel symbols are given in terms of $\partial_c h_{ab}$ by the formula 
\beq
\Gamma_{abc} = \frac{1}{2}(\partial_c h_{ab}+\partial_b h_{ac}-\partial_a h_{bc} ).
\eeq
The linearized Riemann and Weyl tensors are
\bea
&& \hspace{-1cm} R_{abcd}=\frac{1}{2}(\partial_{bc}h_{ad}+\partial_{ad}h_{bc}- \partial_{ac}h_{bd}-\partial_{bd}h_{ac})
= \partial_c\Gamma_{abd} - \partial_d\Gamma_{abc} \\
&& \hspace{-1cm} C_{abcd} = R_{abcd} + \frac{1}{2}(-R_{ac}\eta_{bd} + R_{ad}\eta_{bc} + R_{bc}\eta_{ad} - R_{bd}\eta_{ac})
+\frac{1}{6}(\eta_{ac}\eta_{bd}-\eta_{ad}\eta_{bc})R.
\eea

For expressing $R_{abcd}$, $R_{ab}$ and $R$ in terms of $F_{abc}$,
it is convenient to use $\mathring{F}_{abc}$ (see (\ref{eq.circx}) and (\ref{eq.circ2})). One has
\bea
\label{eq.r1}
&& R_{abcd}  =  \partial_d \mathring{F}_{abc}- \partial_c\mathring{F}_{abd}\\
\label{eq.r2}
&& R_{ab} = \eta^{cd} R_{acbd}  =  - \partial_b \mathring{F}_a - \partial_c \mathring{F}^c{}_{ab}\\
\label{eq.r3}
&& R = \eta^{ab}R_{ab}  =  -2\partial_a\mathring{F}^a  =  \partial_a F^a.
\eea
The linearized Weyl tensor can be expressed in terms of $\breve{F}_{abc}$:
\beq
\label{eq.weyl2}
C_{abcd} = \partial_d \breve{F}_{abc}- \partial_c\breve{F}_{abd}
+\frac{1}{2}( \partial_e \breve{F}^e{}_{ac}\eta_{bd} 
+\partial_e \breve{F}^e{}_{bd}\eta_{ac}
- \partial_e \breve{F}^e{}_{bc}\eta_{ad}
- \partial_e \breve{F}^e{}_{ad}\eta_{bc}).
\eeq
In (\ref{eq.weyl2}) it is not obvious that the right hand side is symmetric with respect to $a,b\leftrightarrow c,d$,
nevertheless (\ref{eq.weyl2}) can be rewritten in a longer form in which this symmetry is obvious:
\bea
&& C_{abcd} = \frac{1}{2}(\partial_d \breve{F}_{abc}- \partial_c\breve{F}_{abd} + \partial_b \breve{F}_{cda}- \partial_a\breve{F}_{cdb}) \nonumber\\
&& \hspace{15mm}+\,\frac{1}{4}( \partial_e \breve{F}^e{}_{ac}\eta_{bd} 
+\partial_e \breve{F}^e{}_{bd}\eta_{ac}
- \partial_e \breve{F}^e{}_{bc}\eta_{ad}
- \partial_e \breve{F}^e{}_{ad}\eta_{bc})\nonumber\\
&& \hspace{15mm}+\,\frac{1}{4}( \partial_e \breve{F}^e{}_{ca}\eta_{bd} 
+\partial_e \breve{F}^e{}_{db}\eta_{ac}
- \partial_e \breve{F}^e{}_{cb}\eta_{ad}
- \partial_e \breve{F}^e{}_{da}\eta_{bc}).
\label{eq.weyl3}
\eea
(\ref{eq.weyl3}) has the form of the Weyl--Lanczos equation for linearized gravity \cite{DK}.
For the linearized Einstein tensor, see (\ref{eq.lg.le}) and (\ref{eq.lg.x3}).

\section{On the gravitational counterpart of the homogeneous Maxwell equations}
\label{app.nn}

In \cite{NN} (see also \cite{NN2}) it is claimed (see Lemma 1 and the subsequent statements in Section 2.1 of \cite{NN})
that if
\beq
\label{eq.nn1}
\partial_c\tilde{F}^{cab}+\partial_c\tilde{F}^{cba}=0,
\eeq
where $F_{abc}$ is assumed to have the algebraic properties (\ref{eq.lg.x1}) and (\ref{eq.lg.x2}),
then there exists
a symmetric tensor potential for $F_{abc}$
(so that the relation between the potential and $F_{abc}$ is given by (\ref{eq.lg.x6})),
and the converse is also true.
It is also stated in \cite{NN} (see equation (5) of \cite{NN}),
that (\ref{eq.nn1}) is equivalent with
\beq
\label{eq.nn2}
\partial_a \mathring{F}_{bcd}+\partial_b \mathring{F}_{cad} + \partial_c \mathring{F}_{abd}=0,
\eeq
where the definition of $\mathring{F}_{abc}$ is understood to be
$\mathring{F}_{abc} = F_{abc} - \frac{1}{2} (F_a\eta_{bc}-F_b\eta_{ac})= F_{abc} - \frac{3}{2}\check{F}_{abc}$
(see (\ref{eq.circ2})).
The derivation of these statements is not given in \cite{NN}.
Our aim in this Appendix is to examine their validity and to correct them.
For simplicity, we assume that $H_{\mathrm{dR}}^2(\Omega)=0$ holds for
the spacetime domain $\Omega$ where $F_{abc}$ is defined.

(\ref{eq.nn2}) is clearly equivalent with $\partial_c \tilde{\mathring{F}}^{cab}=0$, and
in Section \ref{sec.lg.foe} we have seen that the latter equality is indeed necessary and sufficient for
the existence of a potential for $F_{abc}$.

On the other hand, although (\ref{eq.nn1}) follows from (\ref{eq.nn2}), the converse is not true,
therefore (\ref{eq.nn1}) does not imply the existence of a potential for $F_{abc}$.
In order to prove this, let us first note that it is not difficult to verify, using (\ref{eq.trx1}), that
\beq
\label{eq.nn3}
\partial_c \tilde{\check{F}}^{cab} + \partial_c \tilde{\check{F}}^{cba}=0.
\eeq
This identity is a direct consequence of (\ref{eq.trx1}) and holds for any covector field $F_a$.
From (\ref{eq.nn3}) it clearly follows that if $\partial_c \tilde{\mathring{F}}^{cab}=0$, i.e.\ if (\ref{eq.nn2}) holds,
then (\ref{eq.nn1}) also holds.  
Considering the reverse direction, let $V_a$ be an arbitrary covector field and let us define $F_{abc}$ as
$F_{abc} = V_a\eta_{bc} - V_b\eta_{ac}$.
This $F_{abc}$ has the algebraic properties (\ref{eq.lg.x1}) and (\ref{eq.lg.x2}), and it satisfies (\ref{eq.nn1}).
Furthermore, $\mathring{F}_{abc}=-\frac{1}{2}F_{abc}$, thus $\tilde{\mathring{F}}^{abc}=-\frac{1}{2}\tilde{F}^{abc}=
\frac{1}{2}\epsilon^{abcd}V_d$,
and $\partial_a\tilde{\mathring{F}}^{abc} = \frac{1}{2}\epsilon^{abcd}\partial_a V_d$. This is not zero if
$\partial_a V_b-\partial_b V_a\ne 0$, therefore (\ref{eq.nn2}) does not follow from (\ref{eq.nn1}).

\end{document}